\documentclass{elsarticle}

\usepackage[reviewcopy]{adndt}
\usepackage{longtable}

\usepackage{amsmath}

\usepackage{amssymb}

\biboptions{square,sort&compress}
\bibpunct[]{[}{]}{,}{n}{}{;}
\begin{document}

\begin{frontmatter}

\journal{Atomic Data and Nuclear Data Tables}

%% Author, fill in article title here

\title{Discovery of the Isotopes with 11 $\le$ Z $\le$ 19}

%% Fill in author list here
  \author{M. Thoennessen\corref{cor1}}\ead{thoennessen@nscl.msu.edu}

  \cortext[cor1]{Corresponding author.}

  \address{National Superconducting Cyclotron Laboratory and \\ Department of Physics and Astronomy, Michigan State University, \\ East Lansing, MI 48824, USA}

%\date{April 10, 2010} %please do not use \today, use actual date of version

\begin{abstract}
A total of 194 isotopes with 11 $\le$ Z $\le$ 19 have been identified to date. The discovery of these isotopes which includes the observation of unbound nuclei, is discussed. For each isotope a brief summary of the first refereed publication, including the production and identification method, is presented.

\end{abstract}

\end{frontmatter}

%%% Keywords and subject classification are not used in ADNDT
%%%\begin{keywords}
%%%Insert list of keywords here.
%%%\end{keywords}

%%%\begin{subject}[Insert header for classifications]
%%%Use only if your journal has a subject classification requirement
%%%\end{subject}

%%% The table of contents should start a new page. This command will
%%% automatically pull all the unstarred \section, \subsection and
%%% \subsubsection titles into the Contents. Starred versions need to be
%%% done manually using
%%%            \addcontentsline{toc}{[[sub]sub]section}{Section title}
%%% at the correct place. Examples are given below.

%%% The lists of data figures and data tables are created automatically
%%% by the \listofDfigures and \listofDtables commands.

\newpage
\tableofcontents
\listofDfigures
\listofDtables

\vskip5pc

\section{Introduction}\label{s:intro}

The discovery of isotopes of the elements from sodium to potassium is discussed as part of the series of the discovery of isotopes which began with the cerium isotopes in 2009 \cite{2009Gin01}. Guidelines for assigning credit for discovery are (1) clear identification, either through decay-curves and relationships to other known isotopes, particle or $\gamma$-ray spectra, or unique mass and Z-identification, and (2) publication of the discovery in a refereed journal. The authors and year of the first publication, the laboratory where the isotopes were produced as well as the production and identification methods are discussed. When appropriate, references to conference proceedings, internal reports, and theses are included. When a discovery includes a half-life measurement the measured value is compared to the currently adopted value taken from the NUBASE evaluation \cite{2003Aud01} which is based on the ENSDF database \cite{2008ENS01}. In cases where the reported half-life differed significantly from the adopted half-life (up to approximately a factor of two), we searched the subsequent literature for indications that the measurement was erroneous. If that was not the case we credited the authors with the discovery in spite of the inaccurate half-life.

%The first criterium excludes measurements of half-lives of a given element without mass identification. This affects mostly isotopes first observed in fission where decay curves of chemically separated elements were measured without the capability to determine their mass. Also the four-parameter measurements (see, for example, Ref. \cite{1970Joh01}) were, in general, not considered because the mass identification was only $\pm$1 mass unit.

%The second criterium affects especially the isotopes studied within the Manhattan Project. Although an overview of the results was published in 1946 \cite{1946TPP01}, most of the papers were only published in the Plutonium Project Records of the Manhattan Project Technical Series, Vol. 9A, ''Radiochemistry and the Fission Products,'' in three books by Wiley in 1951 \cite{1951Cor01}. We considered this first unclassified publication to be equivalent to a refereed paper. Good examples why publications in conference proceedings should not be considered are $^{118}$Tc and $^{120}$Ru which had been reported as being discovered in a conference proceeding \cite{1996Cza01} but not in the subsequent refereed publication \cite{1997Ber01}.

The initial literature search was performed using the databases ENSDF \cite{2008ENS01} and NSR \cite{2008NSR01} of the National Nuclear Data Center at Brookhaven National Laboratory. These databases are complete and reliable back to the early 1960's. For earlier references, several editions of the Table of Isotopes were used \cite{1940Liv01,1944Sea01,1948Sea01,1953Hol02,1958Str01,1967Led01}. A good reference for the discovery of the stable isotopes was the second edition of Aston's book ``Mass Spectra and Isotopes'' \cite{1942Ast01}.

\section{Discovery of Isotopes with 11 $\le$ Z $\le$ 19}

The discovery of 194 isotopes with 11 $\le$ Z $\le$ 19 includes 19 sodium, 21 magnesium, 22 aluminum, 23 silicon, 21 phosphorus, 22 sulfur, 21 chlorine, 23 argon, and 22 potassium isotopes.

The discovery of the light stable isotopes is not easily defined because they were involved in the discovery of isotopes themselves. We decided to credit discovery if the detection method was sensitive enough to separate isotopes and if isotopes were specifically searched for. Thus the first description of a mass spectrograph by Dempster in 1918 was not considered because Dempster observed only a single isotope per element and did not perform absolute mass measurements \cite{1918Dem01}.

Nuclei beyond the driplines, i.e. nuclei which decay by the emission of a neutron or a proton were included in the compilation. In some cases these nuclei live for only very short times and especially for nuclei beyond the neutron dripline they can only be measured as resonances. Nevertheless the masses can be determined by transfer reactions or by invariant mass measurements and the lifetimes can be determined from the width of the resonance. However, especially for nuclei which are removed by two or more neutrons from the last particle-stable isotope, these resonant states can be very broad and it becomes questionable if the corresponding lifetimes are long enough to be called a nucleus \cite{2004Tho01}. Only unbound nuclei for which a resonance was observed are included, the first ``non-existence'' for an unbound nucleus has been compiled elsewhere \cite{2004Tho01}.

For heavier isotopes we have adopted the practice to accept the observation of isomeric states or excited proton unbound states prior to the determination of the ground states as the discovery. Accordingly for the light nuclei the observed resonance does not necessarily have to correspond to the ground state.

\subsection{Sodium}\vspace{0.0cm}

The observation of 19 sodium isotopes has been reported so far, including 1 stable, 3 proton-rich, 13 neutron-rich, and 2 proton-unbound resonances. The one-neutron unbound resonances of $^{36}$Na and $^{38}$Na should be able to be observed in the future. In addition, $^{39}$Na still might be particle-stable.

\begin{figure}
	\centering
	\includegraphics[scale=0.7]{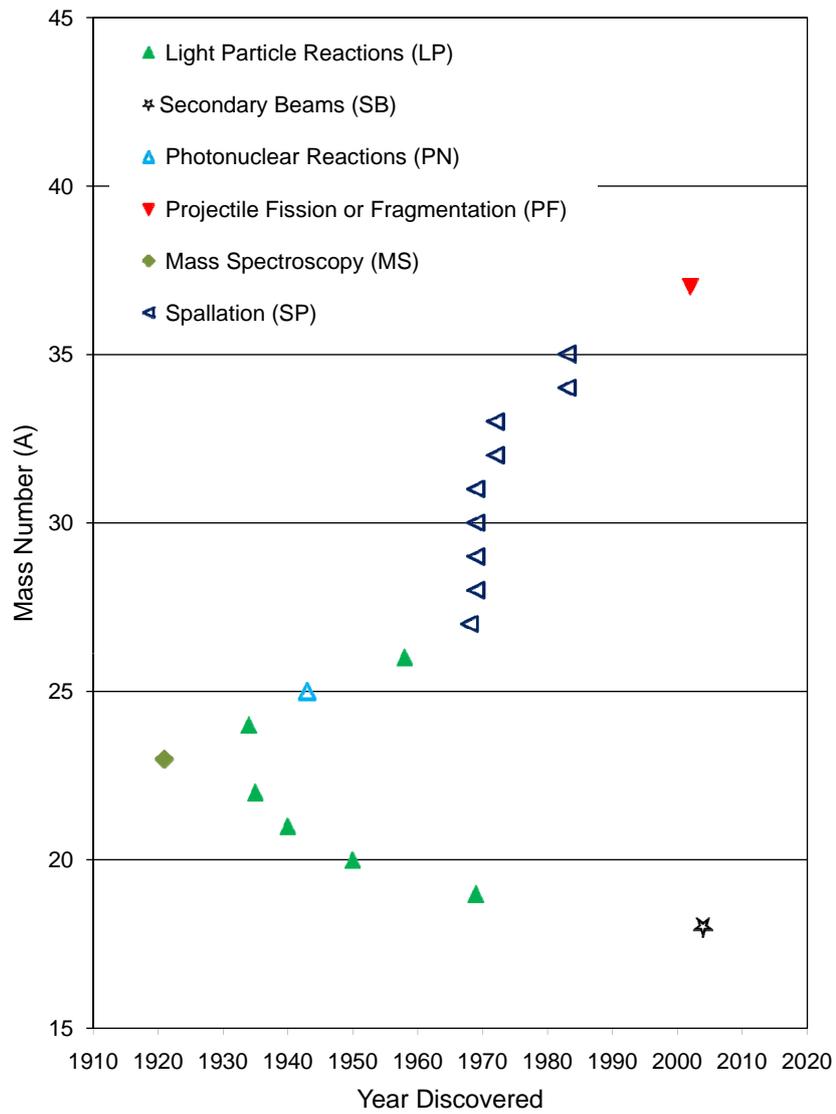}
	\caption{Sodium isotopes as a function of time when they were discovered. The different production methods are indicated.}
\label{f:year-na}
\end{figure}

Figure \ref{f:year-na} summarizes the year of first observation for all sodium isotopes identified by the method of discovery. The radioactive sodium isotopes were produced using light-particle reactions (LP), photo-nuclear reactions (PN), spallation (SP), projectile fragmentation of fission (PF), and most recently with secondary beams (SB). The stable isotopes were identified using mass spectroscopy (MS). Light particles are defined as incident particles with A$\le$4. The discovery of each sodium isotope is discussed in detail and a summary is presented in Table 1.

\subsubsection*{$^{18}$Na}

Zerguerras et al. discovered $^{18}$Na as reported in the 2004 paper ``Study of light proton-rich nuclei by complete kinematics measurements'' \cite{2004Zer01}. A secondary $^{20}$Mg beam was produced by projectile fragmentation with the GANIL SISSI solenoids and ALPHA spectrometer from a primary 95 MeV/nucleon $^{24}$Mg beam. $^{18}$Na was identified from the invariant mass obtained from the $^{17}$Ne and proton events. ``To construct the invariant-mass spectrum of $^{18}$Na, $^{17}$Ne + p events were analysed. The result is shown in [the figure]. A two-peak structure is clearly visible with mass excess values of 24.19(16)~MeV and 25.04(17)~MeV.''

\subsubsection*{$^{19}$Na}

The first observation of $^{19}$Na was described by Cerny et al. ``New nuclides $^{19}$Na and $^{23}$Al observed via the (p,$^6$He) reaction'' in 1969 \cite{1969Cer01}. A $^{24}$Mg target was bombarded with 54.7~MeV protons from the Berkeley 88-in. cyclotron. $^{19}$Na was produced in the reaction $^{24}$Mg(p,$^6$He) and identified by measuring the $^6$He in a two-counter telescope. ``The mass excess of $^{19}$Na is determined to be 12.974$\pm$0.070~MeV [12C=0]... We will take it to be the ground state noting that in either event $^{19}$Na is proton unstable; with this assumption, $^{19}$Na is unbound to $^{18}$Ne + p by 366$\pm$70~keV.''

\subsubsection*{$^{20}$Na}

$^{20}$Na was discovered by Alvarez in the 1950 paper ``Three new delayed alpha-emitters of low mass'' \cite{1950Alv01}. A 32 MeV proton beam from the Berkeley linear accelerator bombarded a proportional counter filled with neon and delayed heavy particles were observed in the counter. ``Two new positron active isotopes, B$^8$ and Na$^{20}$, have been found to decay to excited states of Be$^8$ and Ne$^{20}$, which in turn decay ``instantaneously'' by alpha-emission. Their half-lives are 0.65$\pm$0.1 sec. and 1/4 sec., respectively.'' The half-life of $^{20}$Na is within a factor of two of the accepted value of 447.9(23)~ms.

\subsubsection*{$^{21}$Na}

In the 1940 paper ``Transmutation of the separated isotopes of neon by deuterons'' Pollard et al. reported the observation of $^{21}$Na \cite{1940Pol01}. Neon gas was bombarded with 2.6~MeV deuterons and the isotopes were separated by thermal diffusion. $^{21}$Na produced in the reaction $^{20}$Ne(d,n) and decay curves and absorption spectra were measured. ``A rather weak gamma-ray was found but this did not decay with the 43-second half-life expected. In [the figure] the decay curve is shown. It has a half-life of 26$\pm$3 seconds and is almost certainly to be identified with Na$^{21}$ discovered by Creutz, Fox, and Sutton \cite{1940Cre01} and here produced by the reaction Ne$^{20}$+H$^2\rightarrow$Na$^{21}$+n.'' This half-life agrees with the currently adopted value of 22.49(4)~(s). The work by Creutz et al. mentioned in the quote was only published as an abstract of a meeting.

\subsubsection*{$^{22}$Na}

$^{22}$Na was discovered in 1935 by Frisch as reported in the paper ``Induced radioactivity of fluorine and calcium'' \cite{1935Fri01}. Alpha particles from a 600 mCi radon source were used to irradiate sodium fluoride and lithium fluoride and $^{22}$Na was formed in the reaction $^{19}$F($\alpha$,n). ``The search for Na$^{22}$ was therefore continued with sodium fluoride and lithium fluoride. In both cases weak activity was observed after prolonged bombardment. A chemical separation, kindly carried out by Prof. G. von Hevesy, showed that the active body follows the reactions of sodium, and therefore is presumably Na$^{22}$.'' The estimated half-life between one and several years is consistent with the currently adopted value of 2.6019(4)~y.

\subsubsection*{$^{23}$Na}

The discovery of stable $^{23}$Na was reported by Aston in his 1921 paper ``The constitution of the alkali metals'' \cite{1921Ast03}. The positive anode ray method was used to identify $^{23}$Na with the Cavendish mass spectrograph. ``Sodium (atomic weight 23.00) is the easiest metal to deal with; its mass spectrum consists of a single line only. From the known values of the fields employed this line is in the position expected from the atomic weight; it is therefore assumed to be exactly 23, and used as a standard comparison line.'' Dempster had observed sodium in his mass spectrograph but did not determine the mass independently but rather placed the sodium line to the known chemical mass, he also did not attempt to search for other sodium isotopes \cite{1918Dem01}.

\subsubsection*{$^{24}$Na}

$^{24}$Na was discovered by Fermi et al. in the 1934 article ``Artificial radioactivity produced by neutron bombardment'' \cite{1934Fer01}. Magnesium targets were irradiated with neutrons from a 800 mCi radon beryllium source and activities were measured with Geiger-M\"uller counters following chemical separation. ``The active element decaying with the 15 hours' period could be chemically separated. The irradiated magnesium was dissolved, and a sodium salt was added. The magnesium was then precipitated as phosphate and found to be inactive, while the sodium which remains in the solution carries the activity. The active atom is thus proved not to be an isotope of magnesium, and as neon also can be excluded, we assume it to be an isotope of sodium, formed according to the reaction: Mg$^{24}_{12}$ + n$^1_0$ = Na$^{24}_{11}$ + H$^1_1$.'' This half-life agrees with the currently adopted value of 14.9590(12)~h.

\subsubsection*{$^{25}$Na}

First evidence of $^{25}$Na was shown by Huber et al. in ``Kernphotoeffekt unter Abspaltung eines Protons: Mg$^{26}$($\gamma$,p)Na$^{25}$'' in 1943 \cite{1943Hub02}. Lithium $\gamma$-rays ($\sim$17~MeV) irradiated magnesium targets and the $\beta$-activity and absorption spectra were recorded. ``Die gefundene 62 sec-Aktivit\"at ist somit dem Prozess Mg$^{26}$($\gamma$,p)Na$^{25}$ zuzuschreiben.'' [The observed activity of 62~s is therefore assigned to the process Mg$^{26}$($\gamma$,p)Na$^{25}$]. This half-life agrees with the presently adopted value of 59.1(6)~s. The authors had reported this half-life (62(2)~s) earlier but could not rule out the possibility that it was due to excited states of $^{24}$Mg or $^{25}$Mg \cite{1943Hub01}.

\subsubsection*{$^{26}$Na}

In 1958 Nurmia and Fink observed $^{26}$Na for the first time in ``Cross-sections for (n,p) and (n,$\alpha$) reactions of magnesium with 14.8 MeV neutrons; A new isotope Na$^{26}$'' \cite{1958Nur02}. Neutrons of 14.8 MeV produced at the Arkansas 400~keV Cockroft-Walton accelerator irradiated highly enriched $^{26}$MgO targets. The resulting activities were measured with a plastic scintillator. ``In addition to l5.0-hour Na$^{24}$, a composite activity from a mixture of 38-second Ne$^{23}$ from the Mg$^{26}$(n,$\alpha$) reaction and 60-second Na$^{25}$ from the Mg$^{25}$(n,p) reaction, and an activity with half-life of 1.04$\pm$0.03~s were observed. The last activity, consisting mainly of high energy ($>$5 MeV) beta-particles, has not been reported previously, and it is assigned to a new isotope of sodium Na$^{25}$.'' This half-life agrees with the presently adopted value of 1.077(5)~s.

\subsubsection*{$^{27}$Na}

Klapisch et al. discovered $^{27}$Na in 1968 in``Isotopic distribution of sodium fragments emitted in high-energy nuclear reactions. Identification of $^{27}$Na and possible existence of heavier Na isotopes'' \cite{1968Kla01}. $^{100}$Mo, tantalum, iridium, and uranium targets were bombarded with 10.5~GeV protons from the CERN proton synchrotron and $^{27}$Na was extracted by surface ionization and identified with a Nier-type separator. ``We have found peaks at masses 27, 28, and 29. To the extent that we specifically ionize sodium, this would indicate the existence of three new isotopes of sodium. However, despite their smaller ionization probability, we have to take into account the possibility that Al isotopes produced in the reaction could also be ionized. Thus we see at once that mass 27 cannot be due to Al because the cross section for the production of $^{27}$Al would be some 30 times greater than that for $^{23}$Na, and this is completely ruled out on the basis of nuclear-reaction systematics.''

\subsubsection*{$^{28-31}$Na}

$^{28}$Na, $^{29}$Na, $^{30}$Na, and $^{31}$Na were first identified by Klapisch et al. in the 1969 paper ''Half-Life of $^{11}$Li, of $^{27}$Na, and of the new isotopes $^{28}$Na, $^{29}$Na, $^{30}$Na, and $^{31}$Na produced in high-energy nuclear reactions'' \cite{1969Kla01}. The CERN proton synchrotron was used to bombard iridium and uranium targets with 24 GeV protons. The isotopes were identified with an on-line mass separator and $\beta$ activities were measured with a plastic scintillator. ``Our experimental results establish the particle stability of all the sodium isotopes filling the sd-neutron shell.'' Half-lives of 34(1)~s ($^{28}$Na), 47(3)~s ($^{29}$Na), 55(3)~s ($^{30}$Na), and 16.5(40)~s ($^{31}$Na) were listed in a table and agree with the presently accepted values of 30.5(4)~s, 44.9(12)~s, 48.4(17)~s, and 17.0(4)~s, respectively.

\subsubsection*{$^{32,33}$Na}

In 1972 Klapisch et al. reported the first observation of $^{32}$Na and $^{33}$Na in ``Half-life of the new isotope $^{32}$Na; Observation of $^{33}$Na and other new isotopes produced in the reaction of high-energy protons on U'' \cite{1972Kla01}. Uranium targets were bombarded with 24 GeV protons from the CERN proton synchrotron. $^{32}$Na and $^{33}$Na were identified by on-line mass spectrometry and decay curves were measured. ``[The figure] then shows that peaks do occur for $^{32}$Na in the positions expected from the calibration with the known isotope $^{22}$Na... A search was made for $^{33}$Na using the same procedure during a 3-h experiment with a total of 7$\times$10$^{15}$ protons, and [the table] gives the number of counts at the locations where $^{33}$Na peaks are expected. It is seen that a significant number of counts over the background arises for the first three pairs of peaks. Adding them channel by channel, two peaks of $^{33}$Na are found with 86$\pm$15 and 61$\pm$15 counts, respectively, after a background of 76$\pm$9 has been subtracted.'' Half-lives of 14.5(3)~s ($^{32}$Na) and 20(15)~s ($^{33}$Na) were listed in a table and are consistent with the currently accepted values of 12.9(7)~s and 8.2(2)~s, respectively.

\subsubsection*{$^{34,35}$Na}

Langevin et al. reported the observation of $^{34}$Na and $^{35}$Na in the 1983 paper ``$^{35}$Na: A new neutron-rich sodium isotope'' \cite{1983Lan02}. The CERN synchrotron was used to bombard an iridium target with 10 GeV protons. $^{35}$Na and $^{35}$Na were identified with the on-line mass spectrometer. Beta-delayed neutrons were measured with a NE213 liquid scintillator. ``During the collection of alkali isotopes a multiscaler device defines the time occurrence of each {3-coincident neutron event after each fast extraction beam pulse. [The figure] shows the experimental time occurrence of $\beta$-coincident neutrons for the collection of mass 34 and 35 alkali ions. The 33 events of mass 35 were obtained in 20 h, corresponding to 5$\times$10$^{16}$ protons on the target... A straightforward $\chi^2$ analysis of the experimental results of [the figure] convoluting the time dependence of Na ion production and the $\beta$-decay gives half-lives of (5.5$\pm$1.0) ms for $^{34}$Na and (1.5$\pm$0.5) ms for $^{35}$Na.'' These measured half-lives correspond to the presently adopted values. Langevin et al. did not consider the observation of $^{34}$Na a new discovery referring to earlier conference proceedings and an unpublished Th\`ese de Doctorat. Also, in a 1979 paper D\'etraz et al. showed a figure plotting the number of $^{24}$Na ions per pulse produced by bombarding uranium with 24 GeV protons, however, no further details were given in the paper, stating ``The $\beta$ deccy of $^{34}$Na is certainly accessible to the method...'' \cite{1979Det01}

Langevin et al. reported the discovery of $^{35}$Na in the 1983 paper ``$^{35}$Na: A new neutron-rich sodium isotope'' \cite{1983Lan02}. The CERN synchrotron was used to bombard an iridium target with 10 GeV protons. $^{35}$Na was identified with the on-line mass spectrometer. Beta-delayed neutrons were measured with a NE213 liquid scintillator. ``During the collection of alkali isotopes a multiscaler device defines the time occurrence of each {3-coincident neutron event after each fast extraction beam pulse. [The figure] shows the experimental time occurrence of $\beta$-coincident neutrons for the collection of mass 34 and 35 alkali ions. The 33 events of mass 35 were obtained in 20 h, corresponding to 5$\times$10$^{16}$ protons on the target.'' The measured half-life of 1.5(5)~s corresponds to the presently adopted value.

\subsubsection*{$^{37}$Na}

In the 2002 article ``New neutron-rich isotopes, $^{34}$Ne, $^{37}$Na and $^{43}$Si, produced by fragmentation of a 64A MeV $^{48}$Ca beam'' Notani et al. described the first observation of $^{37}$Na \cite{2002Not01}. The RIKEN ring cyclotron accelerated a $^{48}$Ca beam to 64 MeV/nucleon which was then fragmented on a tantalum target. The projectile fragments were analyzed with the RIPS spectrometer. ``[Part (a) of the figure] shows a two-dimensional plot of A/Z versus Z, obtained from the data accumulated with the $^{40}$Mg B$\rho$ setting, while [part (b)] is for the $^{43}$Si setting. The integrated beam intensities for the two settings are 6.9$\times$10$^{16}$ and 1.7$\times$10$^{15}$ particles, respectively. The numbers of events observed for three new isotopes, $^{34}$Ne, $^{37}$Na and $^{43}$Si, were 2, 3 and 4, respectively.'' Lukyanov et al. reported the discovery of $^{37}$Na independently less than two months later \cite{2002Luk01}.

\subsection{Magnesium}\vspace{0.0cm}

The observation of 21 magnesium isotopes has been reported so far, including 3 stable, 4 proton-rich, 13 neutron-rich, and 1 proton-unbound resonance. The one-neutron unbound resonances of $^{39}$Mg and $^{41}$Mg should be able to be observed in the future. In addition, $^{42}$Mg still might be particle-stable.

\begin{figure}
	\centering
	\includegraphics[scale=0.7]{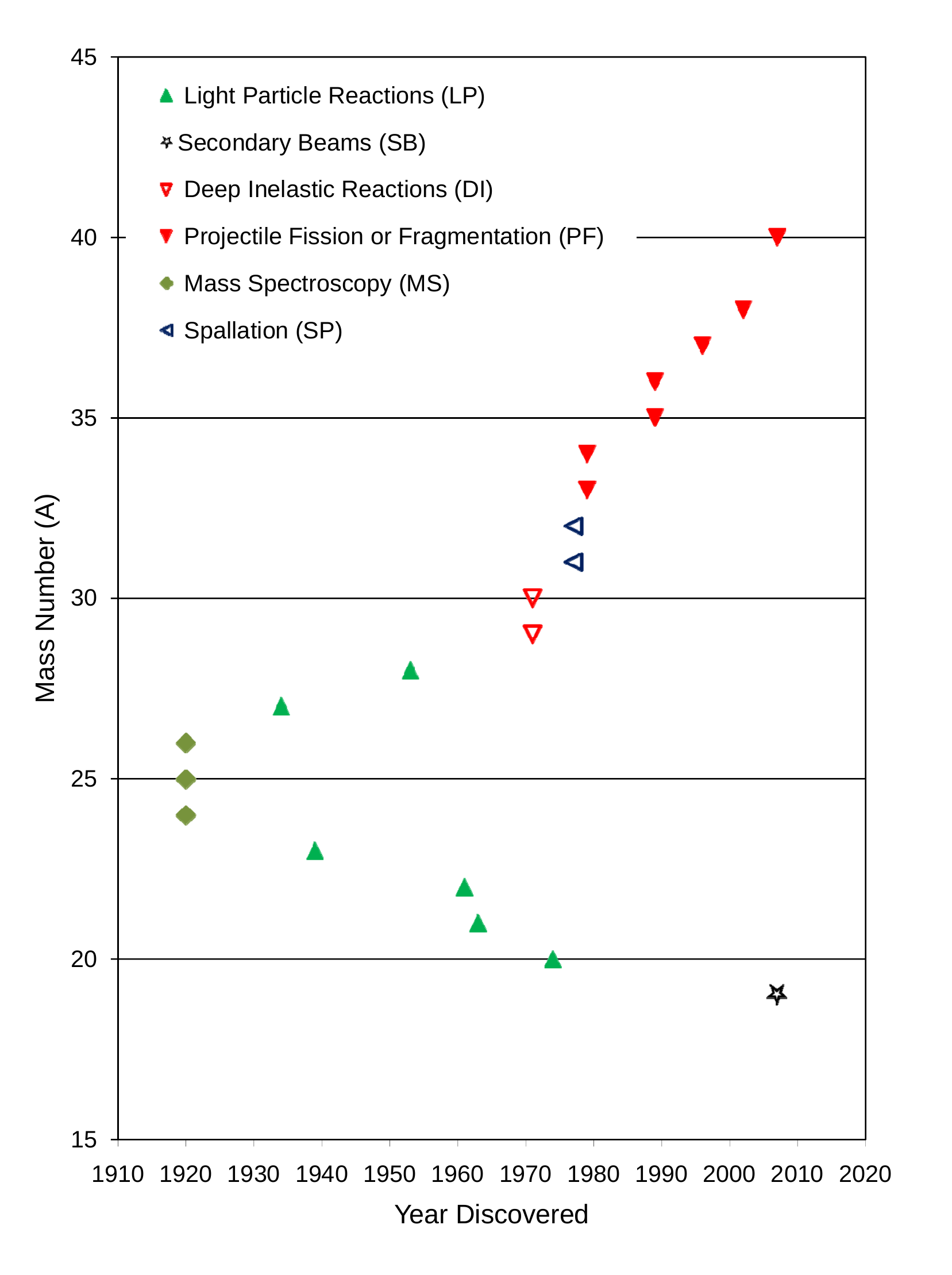}
	\caption{Magnesium isotopes as a function of time when they were discovered. The different production methods are indicated.}
\label{f:year-mg}
\end{figure}

Figure \ref{f:year-mg} summarizes the year of first observation for all magnesium isotopes identified by the method of discovery. The radioactive magnesium isotopes were produced using light-particle reactions (LP), deep-inelastic reactions (DI), spallation (SP), projectile fragmentation of fission (PF), and most recently with secondary beams (SB). The stable isotopes were identified using mass spectroscopy (MS). Light particles are defined as incident particles with A$\le$4. The discovery of each magnesium isotope is discussed in detail and a summary is presented in Table 1.

\subsubsection*{$^{19}$Mg}

$^{19}$Mg was observed by Mukha et al. in the 2007 paper ``Observation of two-proton radioactivity of $^{19}$Mg by tracking the decay products'' \cite{2007Muk01}. A primary 591 MeV/nucleon $^{24}$Mg from the GSI SIS facility was used to produce a secondary beam of $^{20}$Mg. $^{19}$Mg was then produced in a neutron removal reaction and decayed by the emission of two protons in flight. The lifetime was measured by reconstructing the vertex of the decay to $^{17}$Ne and two protons. ``For the first time, the trajectories of the 2p-decay products, $^{17}$Ne+p+p, have been measured by using tracking microstrip detectors which allowed us to reconstruct the 2p-decay vertices and fragment correlations. The half-life of $^{19}$Mg deduced from the measured vertex distribution is 4.0(15) ps in the system of $^{19}$Mg. The Q value of the 2p decay of the $^{19}$Mg ground state inferred from the measured p$-$p$-^{17}$Ne correlations is 0.75(5) MeV.'' This is currently the only measured half-life.

\subsubsection*{$^{20}$Mg}

$^{20}$Mg was discovered by Robertson et al. in the 1974 paper ``Highly proton-rich T$_z$ = $-$2 nuclides: $^8$C and $^{20}$Mg'' \cite{1974Rob01}. Alpha-particles accelerated to 156 MeV by the J\"ulich isochronous cyclotron bombarded an enriched $^{24}$Mg target and produced $^{20}$Mg in the reaction $^{24}$Mg($\alpha$,$^8$He). The $^8$He ejectiles were measured in a double-focusing magnetic analyzer and the energy-loss, energy, magnetic rigidity and time-of-flight were recorded. ``For $^{20}$Mg, a mass excess of 17.74$\pm$0.21 MeV is found, indicating that $^{20}$Mg is nucleon stable.''
An earlier tentative report of a 0.62~s half-life for $^{20}$Mg \cite{1964Mac03} was evidently incorrect.

\subsubsection*{$^{21}$Mg}

In ``Observation of delayed proton radioactivity'' Barton et al. implied the observation of $^{21}$Mg for the first time in 1963 \cite{1963Bar01}. The McGill Synchrocyclotron accelerated protons to 97 MeV which bombarded a magnesium target. $^{21}$Mg was identified by the observation of $\beta$-delayed protons in a silicon junction particle detector. ``The hypothesis that the decay of the nuclide (2k+2,2k$-$1) will dominate the delayed proton spectrum from targets of both element (2k+2) and element (2k+1) seems to be verified. In particular, by following very reasonable rules for predicting proton lines, all the observed lines are accounted for and all those predictions based on known level properties are borne out. The existence of Mg$^{21}$, Ne$^{17}$, and O$^{13}$ is assumed since it seems fairly certain that proton lines in the decay of each have been observed.'' The half-life was subsequently measured by McPherson et al. \cite{1965McP01} who acknowledged the tentative observation by Barton et al.

\subsubsection*{$^{22}$Mg}

The first observation of $^{22}$Mg was reported by Ajzenberg-Selove et al. in the 1961 paper `` Energy levels of Na$^{21}$ and Mg$^{22}$'' \cite{1961Ajz01}. $^3$He ions were accelerated to 4$-$5~MeV by the Los Alamos vertical Van de Graaff generator and bombarded a neon target. $^{22}$Mg was produced in the reaction $^{20}$Ne($^3$He,n) and identified by measuring neutrons with a NE102 scintillator. ``Mg$^{22}$, here reported for the first time, has a mass excess (M$-$A) of $-$0.14$\pm$0.08 Mev (C$^{12}$ reference) from Q = $-$0.043$\pm$0.08 Mev for Ne$^{20}$(He$^3$,n)Mg$^{22}$.'' Previously a 130~ms activity was assigned to either $^{23}$Al or $^{22}$Mg \cite{1954Tyr01}.

\subsubsection*{$^{23}$Mg}

White et al. described the discovery of $^{23}$Mg in 1939 in ``Short-lived radioactivities induced in fluorine, sodium and magnesium by high energy protons'' \cite{1939Whi01}. Sodium chloride targets were bombarded with 6 MeV protons and $^{23}$Mg was produced in the charge-exchange reaction $^{23}$Na(p,n). The resulting activities were measured with a Lauritsen electroscope. ``Sodium, in the form of NaCl, was bombarded for one minute with 6.0-Mev protons. A characteristic decay curve is shown in [the figure]. The half-life, as found by averaging several runs, is 11.6$\pm$0.5~sec. We assume that the activity indicates the production of Mg$^{23}$, for all other possible reactions lead either to stable isotopes or well-known long periods.'' This half-life agrees with the presently accepted value of 11.317(11)~s.

\subsubsection*{$^{24-26}$Mg}

In 1920, Dempster discovered $^{24}$Mg, $^{25}$Mg, and $^{26}$Mg as reported in ``Positive ray analysis of magnesium'' \cite{1920Dem01}. In an adaptation of the positive ray method magnesium was analyzed. ``Using the apparatus for positive ray analysis described in The Physical Review for April, 1918, I have recently succeeded in analyzing the element magnesium (atomic weight 24.36) into three isotopes of atomic weights 24, 25 and 26.''

\subsubsection*{$^{27}$Mg}

$^{27}$Mg was discovered by Fermi et al. in the 1934 article ``Artificial radioactivity produced by neutron bombardment'' \cite{1934Fer01}. Aluminum targets were irradiated with neutrons from a 800 mCi radon beryllium source and activities were measured with Geiger-M\"uller counters. ``13$-$Aluminium: This element acquires a strong activity under neutron bombardment. The decay curves indicate two periods of about 12 minutes (i = 0.8) and 15 hours (i = 0.5)... The active product with the 12-minute period has not been separated. However, we consider it likely to be Mg$^{27}$, as the other two possible cases, Al$^{28}$ and A1$^{26}$, are probably to be excluded, the first because Al$^{28}$, as we shall next see, is a radioactive isotope with a period of 3 minutes, and the latter because Al$^{26}$ should probably disintegrate with emission of positrons.'' This half-life is consistent with the currently adopted value of 9.458(12)~min.

\subsubsection*{$^{28}$Mg}

Sheline and Johnson identified $^{28}$Mg for the first time in 1953 in ``New long-lived magnesium-28 isotope'' \cite{1953She01}. $^{28}$Mg was produced in the reaction Si$^{30}$($\gamma$,2p) with 100~MeV $\gamma$-rays at the Chicago betatron and Mg$^{26}$($\alpha$,2p)Mg$^{28}$ with 39~MeV $\alpha$ particles from the Berkeley 60-in. cyclotron. Activities were measured following chemical separation. ``Magnesium$-$28, a 21=hr. $\beta^-$ emitter, has been produced in both a betatron irradiation and a cyclotron bombardment.'' This half-life agrees with the presently adopted value of 20.915(9)~h.

\subsubsection*{$^{29,30}$Mg}

Artukh et al. discovered $^{29}$Mg and $^{30}$Mg in the 1971 paper ``New isotopes $^{29,30}$Mg, $^{31,32,33}$Al, $^{33,34,35,36}$Si, $^{35,36,37,38}$P, $^{39,40}$S, and $^{41,42}$Cl produced in bombardment of a $^{232}$Th target with 290 MeV $^{40}$Ar ions'' \cite{1971Art01}. A 290 MeV $^{40}$Ar beam from the Dubna 310 cm heavy-ion cyclotron bombarded a metallic $^{232}$Th. Reaction products were separated and identified with a magnetic spectrometer and a surface barrier silicon telescope. ``Apart from the nucleides already known, 17 new nucleides, namely: $^{29,30}$Mg, $^{31,32,33}$Al, $^{33,34,35,36}$Si, $^{35,36,37, 38}$P, $^{39,40}$S and $^{41,42}$Cl have been reliably detected.''

\subsubsection*{$^{31,32}$Mg}

$^{31}$Mg and $^{32}$Mg was discovered by Butler et al. in ``Observation of the new nuclides $^{27}$Ne, $^{31}$Mg, $^{32}$Mg, $^{34}$Al, and $^{39}$P'' in 1977 \cite{1977But01}. $^{31}$Mg and $^{32}$Mg were produced in the spallation reaction of 800 MeV protons from the Clinton P. Anderson Meson Physics Facility LAMPF on a uranium target. The spallation fragments were identified with a silicon $\Delta$E-E telescope and by time-of-flight measurements. ``All of the stable and known neutron-rich nuclides (except $^{24}$O and the more neutron-rich Na isotopes) are seen. The five previously unobserved neutron-rich nuclides $^{27}$Ne, $^{31}$Mg, $^{32}$Mg, $^{34}$Al, and $^{39}$P are clearly evident. Each of these peaks contains ten or more events.''

\subsubsection*{$^{33,34}$Mg}

The first observation of $^{33}$Mg and $^{34}$Mg was reported by Westfall et al. in ``Production of neutron-rich nuclides by fragmentation of 212-MeV/amu $^{48}$Ca'' in 1979 \cite{1979Wes01}. $^{48}$Ca ions (212 MeV/nucleon) from the Berkeley Bevalac were fragmented on a beryllium target. The fragments were selected by a zero degree spectrometer and identified in a telescope consisting of 12 Si(Li) detectors, 2 position-sensitive Si(Li) detectors, and a veto scintillator. ``In this letter, we present the first experimental evidence for the particle stability of fourteen nuclides $^{22}$N, $^{26}$F, $^{33,34}$Mg, $^{36,37}$Al, $^{38,39}$Si, $^{41,42}$P, $^{43,44}$S, and $^{44,45}$Cl produced in the fragmentation of 212-MeV/amu $^{48}$Ca.''

\subsubsection*{$^{35,36}$Mg}

Guillemaud-Mueller et al. announced the discovery of $^{35}$Mg and $^{36}$Mg in the 1989 article ``Observation of new neutron rich nuclei $^{29}$F, $^{35,36}$Mg, $^{38,39}$Al, $^{40,41}$Si, $^{43,44}$P, $^{45-47}$S, $^{46-49}$Cl, and $^{49-51}$Ar from the interaction of 55 MeV/u $^{48}$Ca+Ta'' \cite{1989Gui01}. A 55~MeV/u $^{48}$Ca beam was fragmented on a tantalum target at GANIL and the projectile-like fragments were separated by the zero degree doubly achromatic LISE spetrometer. ``[The figure] represents the two-dimensional plot (energy loss versus time-of-flight) obtained under these conditions after 40~h integration time with an average intensity of 150 enA. The new species $^{35,36}$Mg, $^{38,39}$Al, $^{40,41}$Si, $^{43,44}$P, $^{45,46,47}$S, $^{46,47,48,49}$Cl, and $^{49,50,51}$Ar are clearly visible.''

\subsubsection*{$^{37}$Mg}

$^{37}$Mg was discovered by Sakurai et al. in 1996 as reported in ``Production and identification of new neutron-rich nuclei, $^{31}$Ne and $^{37}$Mg, in the reaction 80A MeV $^{50}$Ti + $^{181}$Ta'' \cite{1996Sak01}. A $^{50}$Ti beam was accelerated at the RIKEN Ring Cyclotron to 80 MeV/nucleon and fragmented on a tantalum target. The fragments were analyzed by the RIPS spectrometer and identified on the basis of energy loss, total kinetic energy, time-of-flight and magnetic rigidity. ``All of the fragments of $^{30,31,32}$Ne, $^{32,33,34,35}$Na, and $^{35,36,37}$Mg were stopped at the SSD4 with the selected window of the magnetic rigidity. Significant numbers of events have been observed for new isotopes, $^{31}$Ne (23 events) and $^{37}$Mg (three events).''

\subsubsection*{$^{38}$Mg}

In the 2002 article ``New neutron-rich isotopes, $^{34}$Ne, $^{37}$Na and $^{43}$Si, produced by fragmentation of a 64A MeV $^{48}$Ca beam'' Notani et al. observed $^{38}$Mg \cite{2002Not01}. The RIKEN ring cyclotron accelerated a $^{48}$Ca beam to 64 MeV/nucleon which was then fragmented on a tantalum target. The projectile fragments were analyzed with the RIPS spectrometer. The observation of $^{38}$Mg was not explicitly mentioned because its discovery was attributed to a previous publication of a conference proceeding \cite{1997Sak01}. In the two-dimensional A/Z versus Z plot for the $^{40}$Mg B$\rho$ setting events for $^{38}$Mg can clearly be identified. It represents the first publication of $^{38}$Mg in a refereed publication.

\subsubsection*{$^{40}$Mg}

Baumann et al. discovered $^{40}$Mg in the 2007 paper ``Discovery of $^{40}$Mg and $^{42}$Al suggests neutron drip-line slant towards heavier isotopes'' \cite{2007Bau01}. A 141 MeV/nucleon $^{48}$Ca beam bombarded a natural tungsten target and $^{40}$Mg was identified with the MSU/NSCL A1900 fragment separator and the S800 analysis system. ``The particle identification can be seen in [the figure], where the locus of isotopes with constant N=2Z is indicated by the vertical line and heavier isotopes lie to the right. Three events of $^{40}$Mg were clearly identified. Each of the parameters that are used for the particle identification has been checked on an event-by-event basis to exclude possible ambiguous background events.''

\subsection{Aluminum}\vspace{0.0cm}

The observation of 22 aluminum isotopes has been reported so far, including 1 stable, 5 proton-rich, and 16 neutron-rich isotopes. The proton dripline has been reached with the observation that $^{21}$Al is unbound \cite{1987Sai01}. According to the HFB-14 model \cite{2007Gor01}, three more aluminum neutron-rich isotopes could be bound ($^{44}$Al, $^{45}$Al, and $^{47}$Al).

\begin{figure}
	\centering
	\includegraphics[scale=0.7]{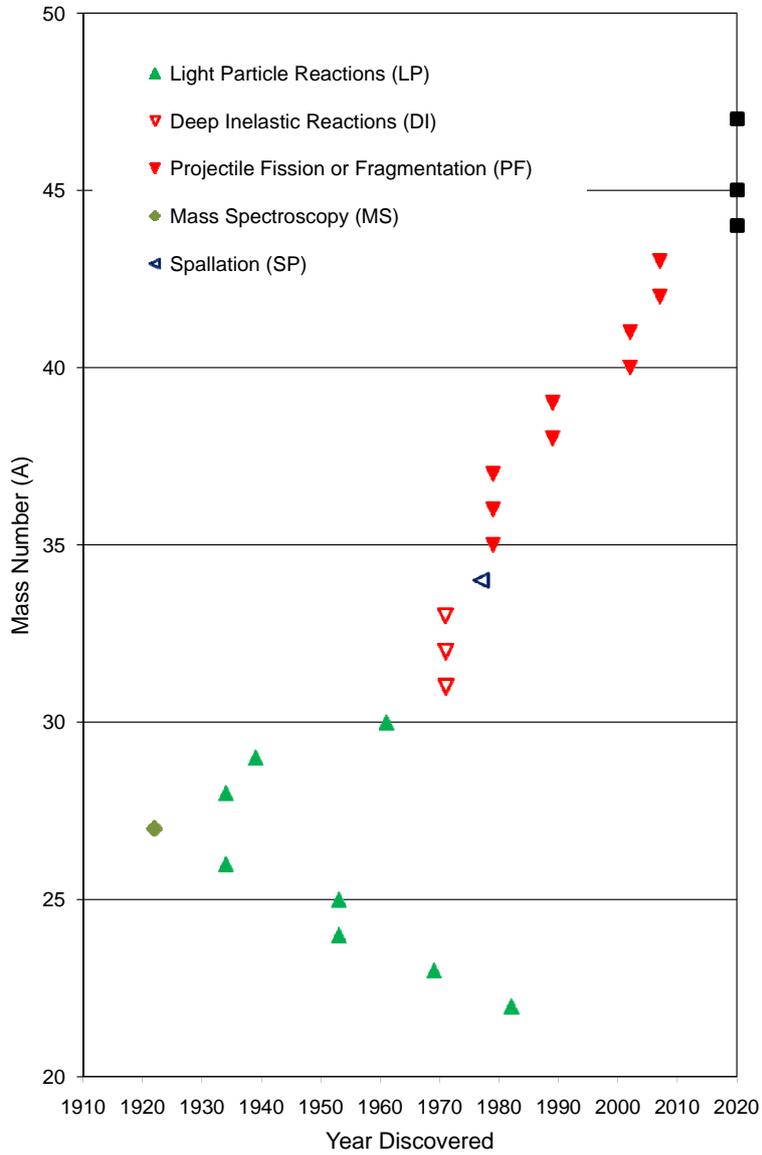}
	\caption{Aluminum isotopes as a function of time when they were discovered. The different production methods are indicated. The solid black squares on the right hand side of the plot are isotopes predicted to be bound by the HFB-14 model.}
\label{f:year-al}
\end{figure}

Figure \ref{f:year-al} summarizes the year of first observation for all aluminum isotopes identified by the method of discovery. The radioactive aluminum isotopes were produced using light-particle reactions (LP), deep-inelastic reactions (DI), spallation (SP), and projectile fragmentation of fission (PF). The stable isotopes were identified using mass spectroscopy (MS). Light particles are defined as incident particles with A$\le$4. The discovery of each aluminum isotope is discussed in detail and a summary is presented in Table 1.

\subsubsection*{$^{22}$Al}

Cable et al. discovered $^{22}$Al as described in the 1982 paper ``Beta-delayed proton decay of an odd-odd T$_z$=$-$2 isotope, $^{22}$Al'' \cite{1982Cab01}. A 110 MeV $^3$He beam from the Berkeley 88-in cyclotron bombarded magnesium targets forming $^{22}$Al in the reaction $^{24}$Mg($^3$He,p4n). Beta-delayed protons were measured with a three-element semiconductor particle telescope. ``At 110 MeV, two new proton groups are observed at laboratory energies of 7.839$\pm$0.015~MeV and 8.149$\pm$0.021~MeV. As noted below and as indicated in [the figure], these groups lie very near the predicted absolute
proton energies for the decay of $^{22}$Al based on Coulomb displacement energy calculations... A rough half-life (70$^{+50}_{-35}$~ms) for the 7.839 group was determined by observing the relative yields of $^{25}$Si, $^{21}$Mg, and $^{22}$Al with different helium jet operating conditions resulting in different transit times from target to catcher.'' This half-life agrees with the presently adopted value of 59(3)~ms.

\subsubsection*{$^{23}$Al}

The first observation of $^{23}$Al was described by Cerny et al. ``New nuclides $^{19}$Na and $^{23}$Al observed via the (p,$^6$He) reaction'' in 1969 \cite{1969Cer01}. A natural silicon target was bombarded with 54.7~MeV protons from the Berkeley 88-in. cyclotron. $^{23}$Al was produced in the reaction $^{28}$Si(p,$^6$He) and identified by measuring the $^6$He in a two counter telescope. ``The mass excess of $^{23}$Al is determined to be 6.766$\pm$0.080~MeV. (The data in [the figure] also show the presence of $^7$B ground state from reactions on a $^{12}$C target impurity.) Therefore, $^{23}$Al is bound to $^{22}$Mg+p by 146$\pm$82~keV and is nucleon stable.'' Previously a 130~ms activity was assigned to either $^{23}$Al or $^{22}$Mg \cite{1954Tyr01}.

\subsubsection*{$^{24}$Al}

In 1953 $^{24}$Al was reported by Glass et al. in ``The short-lived radioisotopes P$^{28}$ and Cl$^{32}$'' \cite{1953Gla01}. Protons were accelerated to 20 MeV by the UCLA cyclotron and bombarded magnesium targets. $^{24}$Al was produced in (p,n) charge exchange reactions and identified by measuring $\gamma$-rays with a NaI crystal. ``We have also obtained some results on Al$^{24}$ from the reaction Mg$^{24}$(p,n)Al$^{24}$. We observe gamma-radiations of energy 7.1$\pm$0.2~Mev, 5.3$\pm$0.2~Mev, 4.3$\pm$0.2~Mev, and 2.9$\pm$0.2~Mev. Our value for the half-life is 2.10$\pm$0.04 seconds which agrees within experimental error with the value obtained by Birge.'' This half-life is in agreement with the currently adopted value of 2.053(4)~s. Birge had reported a half-life of 2.3(2)~s but it was only published as an abstract of a meeting \cite{1953Bir01}.

\subsubsection*{$^{25}$Al}

Churchill et al. reported the discovery of $^{25}$Al in the 1953 paper ``Half-value periods for the decay of aluminium-26, aluminium-25 and nitrogen-13'' \cite{1953Chu01}. Enriched $^{24}$Mg targets were bombarded with 418~keV protons and $^{25}$Al was formed by resonant proton capture. Positron activities were measured with a Geiger-M\"uller tube. ``The half-value periods were calculated from these results using the rigorous treatment given by Peierls and were as follows: aluminium-26, 6.68$\pm$0.11~sec.; aluminium-25, 7.62$\pm$0.13~sec.; and nitrogen-13, 602.9$\pm$1.9~sec.'' The half-life for $^{25}$Al agrees with the currently adopted value of 7.183(12)s. A previously measured half-life of 7.3~s was only published in a meeting abstract \cite{1948Bra01}.

\subsubsection*{$^{26}$Al}

The first observation of $^{26}$Al was reported by Frisch et al. in ``Induced radioactivity of sodium and phosphorus'' in 1934 \cite{1934Fri01}. A 1 mCi thorium B + C $\alpha$ source was used to irradiate sodium targets and the subsequent activity was measured with a Geiger-M\"uller counter. ``Three different sodium compounds (NaCI, NaF, Na$_2$C$_2$O$_4$) have been investigated; they all showed a fairly strong activity, dying off very quickly. The half value period has been determined by recording the impulses on a rotating drum, the whole decay curve being recorded 21 times. The half value period was found to be 7$\pm$1 seconds... So for sodium and phosphorus the reactions would be $_{11}$Na$^{23}$ + $\alpha$ = $_{13}$Al$^{26}$ + neutron and $_{15}$P$^{31}$ + $\alpha$ = $_{17}$Cl$^{34}$ + neutron, respectively.'' This half-life agrees with the presently accepted value of 6.3452(19)~s for the isomeric state.

\subsubsection*{$^{27}$Al}

Aston identified $^{27}$Al in 1922 as reported in ``The isotopes of selenium and some other elements'' \cite{1922Ast03}.  Aluminum was measured with the Cavendish mass spectrometer. ``Application of the method to cadmium and tellurium has failed to give the mass lines of these elements. The employment of the more volatile TeCl$_3$ was also unsuccessful, but incidentally gave evidence of great value, which practically confirms two facts previously suspected, namely, that chlorine has no isotope of mass 39, and that aluminium is a simple element 27.''

\subsubsection*{$^{28}$Al}

Curie and Joliot discovered $^{28}$Al in 1934 in ``I. Production articicielle d'\'el\'ements radioactifs II. Preuve chimique de la transmutation des \'el\'ements.'' \cite{1934Cur02}. Magnesium samples were irradiated by polonium $\alpha$-particles and their electron and positron activities were measured as a function of time. ``Le radio\'el\'ement \'emetteur de rayons $\beta$ cr\'e\'e dans le magn\'esium irradi\'e est probablement un noyau $^{28}_{13}$Al, form\'e \`a partir de $^{25}_{12}$Mg par capture de la particule $\alpha$ et \'emission d'un proton. Les \'electrons n\'egatifs \'etant plus nombreux que les positifs, il est probable que la p\'eriode de 2 mn 15 s observ\'ee correspond a ce radio\'el\'ement.'' [The $\beta$-emitter produced in the irradiation of magnesium is probably the nucleus $^{28}_{12}$Al formed by $\alpha$ capture on $^{25}_{12}$Mg and the emission of a proton. Because there are more negative electrons than positive, it is probable that the observed 2 min 15 s half-life corresponds to this radio-element.] This half-life agrees with the currently adopted value of 2.2414(12)~min.

\subsubsection*{$^{29}$Al}

Bethe and Henderson correctly identified $^{29}$Al in the 1939 paper ``Evidence for incorrect assignment of the supposed Si$^{27}$ radioactivity of 6.7-minute half-life'' \cite{1939Bet01}. The Purdue cyclotron was used to irradiate magnesium with 16 MeV$\alpha$-particles. Photographs were taken with a Wilson cloud chamber after the irradiation. Previously there had been some uncertainties about the assignment of a 6$-$7~min half-life to either $^{27}$Si or $^{29}$Al. ``We have therefore repeated the experiment with the 16-Mev $\alpha$-particles furnished by the Purdue cyclotron. The result is that there is no positron activity, but only negative electrons. This proves that the previous assignment was incorrect and that the 6.7-min. period is almost certainly Al$^{29}$, formed by the reaction Mg$^{26}$($\alpha$,p).'' This half-life agrees with the presently accepted value of 6.56(6)~min. The previous measurements with the uncertain assignment were 7.5(15)~min \cite{1935Eck01}, 6.7(10)~min \cite{1935Fah01}, 6$-$7~min \cite{1936Ell01}, and 6.6(3)~min \cite{1937Mey01}.

\subsubsection*{$^{30}$Al}

Robinson and Johnson discovered $^{30}$Al in 1961 in ``New isotope, Al$^{30}$'' \cite{1961Rob01}. Silicon targets were irradiated with fast neutrons produced by bombarding lithium with 8.8 MeV deuterons from the Purdue 37-in cyclotron. $^{30}$Al was formed in the (n,p) charge exchange reaction and identified by measuring $\gamma$- and $\beta$-rays in a NaI(Tl) crystal and a plastic phosphor detector, respectively. ``A radioisotope with a (3.27$\pm$0.20)-sec half-life is produced by bombarding silicon with (Li$^7$-d) neutrons. This activity is that of the previously unobserved isotope Al$^{30}$ and is produced by the reaction Si$^{30}$(n,p)Al$^{30}$.'' This half-life agrees with the presently adopted value of 3.60(6)~s.

\subsubsection*{$^{31-33}$Al}

Artukh et al. discovered $^{31}$Al, $^{32}$Al, and $^{33}$Al in the 1971 paper ``New isotopes $^{29,30}$Mg, $^{31,32,33}$Al, $^{33,34,35,36}$Si, $^{35,36,37,38}$P, $^{39,40}$S, and $^{41,42}$Cl produced in bombardment of a $^{232}$Th target with 290 MeV $^{40}$Ar ions'' \cite{1971Art01}. A 290 MeV $^{40}$Ar beam from the Dubna 310 cm heavy-ion cyclotron bombarded a metallic $^{232}$Th. Reaction products were separated and identified with a magnetic spectrometer and a surface barrier silicon telescope. ``Apart from the nucleides already known, 17 new nucleides, namely: $^{29,30}$Mg, $^{31,32,33}$Al, $^{33,34,35,36}$Si, $^{35,36,37, 38}$P, $^{39,40}$S and $^{41,42}$Cl have been reliably detected.''

\subsubsection*{$^{34}$Al}

$^{34}$Al was discovered by Butler et al. in ``Observation of the new nuclides $^{27}$Ne, $^{31}$Mg, $^{32}$Mg, $^{34}$Al, and $^{39}$P'' in 1977 \cite{1977But01}. $^{34}$Al was produced in the spallation reaction of 800 MeV protons from the Clinton P. Anderson Meson Physics Facility LAMPF on a uranium target. The spallation fragments were identified with a silicon $\Delta$E-E telescope and by time-of-flight measurements. ``All of the stable and known neutron-rich nuclides (except $^{24}$O and the more neutron-rich Na isotopes) are seen. The five previously unobserved neutron-rich nuclides $^{27}$Ne, $^{31}$Mg, $^{32}$Mg, $^{34}$Al, and $^{39}$P are clearly evident. Each of these peaks contains ten or more events.''

\subsubsection*{$^{35}$Al}

In 1979 Symons et al. described the discovery of $^{35}$Al in ``Observation of new neutron-rich isotopes by fragmentation of 205-MeV/Nucleon $^{40}$Ar ions'' \cite{1979Sym01}. A 205 MeV/nucleon $^{40}$Ar beam from the Berkeley Bevalac was fragmented on a carbon target. The projectile fragments were analyzed with a zero-degree magnetic spectrometer and detected in two detector telescopes. ``Projected mass spectra with a gate of $\pm$0.2 units about charges 10, 11, 12, and 13 are shown in [the figure]. $^{28}$Ne and $^{35}$Al are positively identified as particle-stable isotopes with more than 10 counts in each case.''

\subsubsection*{$^{36,37}$Al}

The first observation of $^{36}$Al and $^{37}$Al was reported by Westfall et al. in ``Production of neutron-rich nuclides by fragmentation of 212-MeV/amu $^{48}$Ca'' in 1979 \cite{1979Wes01}. $^{48}$Ca ions (212 MeV/nucleon) from the Berkeley Bevalac were fragmented on a beryllium target. The fragments were selected by a zero degree spectrometer and identified in a telescope consisting of 12 Si(Li) detectors, 2 position-sensitive Si(Li) detectors, and a veto scintillator. ``In this letter, we present the first experimental evidence for the particle stability of fourteen nuclides $^{22}$N, $^{26}$F, $^{33,34}$Mg, $^{36,37}$Al, $^{38,39}$Si, $^{41,42}$P, $^{43,44}$S, and $^{44,45}$Cl produced in the fragmentation of 212-MeV/amu $^{48}$Ca.''

\subsubsection*{$^{38,39}$Al}

Guillemaud-Mueller et al. announced the discovery of $^{38}$Al and $^{39}$Al in the 1989 article ``Observation of new neutron rich nuclei $^{29}$F, $^{35,36}$Mg, $^{38,39}$Al, $^{40,41}$Si, $^{43,44}$P, $^{45-47}$S, $^{46-49}$Cl, and $^{49-51}$Ar from the interaction of 55 MeV/u $^{48}$Ca+Ta'' \cite{1989Gui01}. A 55~MeV/u $^{48}$Ca beam was fragmented on a tantalum target at GANIL and the projectile-like fragments were separated by the zero degree doubly achromatic LISE spetrometer. ``[The figure] represents the two-dimensional plot (energy loss versus time-of-flight) obtained under these conditions after 40~h integration time with an average intensity of 150 enA. The new species $^{35,36}$Mg, $^{38,39}$Al, $^{40,41}$Si, $^{43,44}$P, $^{45,46,47}$S, $^{46,47,48,49}$Cl, and $^{49,50,51}$Ar are clearly visible.''

\subsubsection*{$^{40,41}$Al}

In the 2002 article ``New neutron-rich isotopes, $^{34}$Ne, $^{37}$Na and $^{43}$Si, produced by fragmentation of a 64A MeV $^{48}$Ca beam'' Notani et al. observed $^{40}$Al and $^{41}$Al \cite{2002Not01}. The RIKEN ring cyclotron accelerated a $^{48}$Ca beam to 64 MeV/nucleon which was then fragmented on a tantalum target. The projectile fragments were analyzed with the RIPS spectrometer. The observation of $^{40}$Al and $^{41}$Al was not explicitly mentioned because its discovery was attributed to a previous publication of a conference proceeding \cite{1997Sak01}. In the two-dimensional A/Z versus Z plot for the $^{40}$Mg B$\rho$ setting events for $^{40}$Al and $^{41}$Al can clearly be identified. It represents the first publication of $^{40}$Al and $^{41}$Al in a refereed publication.

\subsubsection*{$^{42,43}$Al}

Baumann et al. observed $^{42}$Al and $^{43}$Al in the 2007 paper ``Discovery of $^{40}$Mg and $^{42}$Al suggests neutron drip-line slant towards heavier isotopes'' \cite{2007Bau01}. A 141 MeV/nucleon $^{48}$Ca beam bombarded a natural tungsten target and $^{40}$Mg was identified with the MSU/NSCL A1900 fragment separator and the S800 analysis system. ``Further, the 23 events of $^{42}$Al establish its discovery. [The figure] also contains one event consistent with $^{43}$Al.''

\subsection{Silicon}\vspace{0.0cm}

The observation of 23 silicon isotopes has been reported so far, including 3 stable, 6 proton-rich, and 14 neutron-rich isotopes. No specific searches for the existence of $^{21}$Si have been reported and so it could potentially still be observed \cite{2004Tho01}. According to the HFB-14 model \cite{2007Gor01}, three more silicon neutron-rich isotopes could be bound ($^{46}$Si, $^{48}$Si, and $^{50}$Si).

\begin{figure}
	\centering
	\includegraphics[scale=0.7]{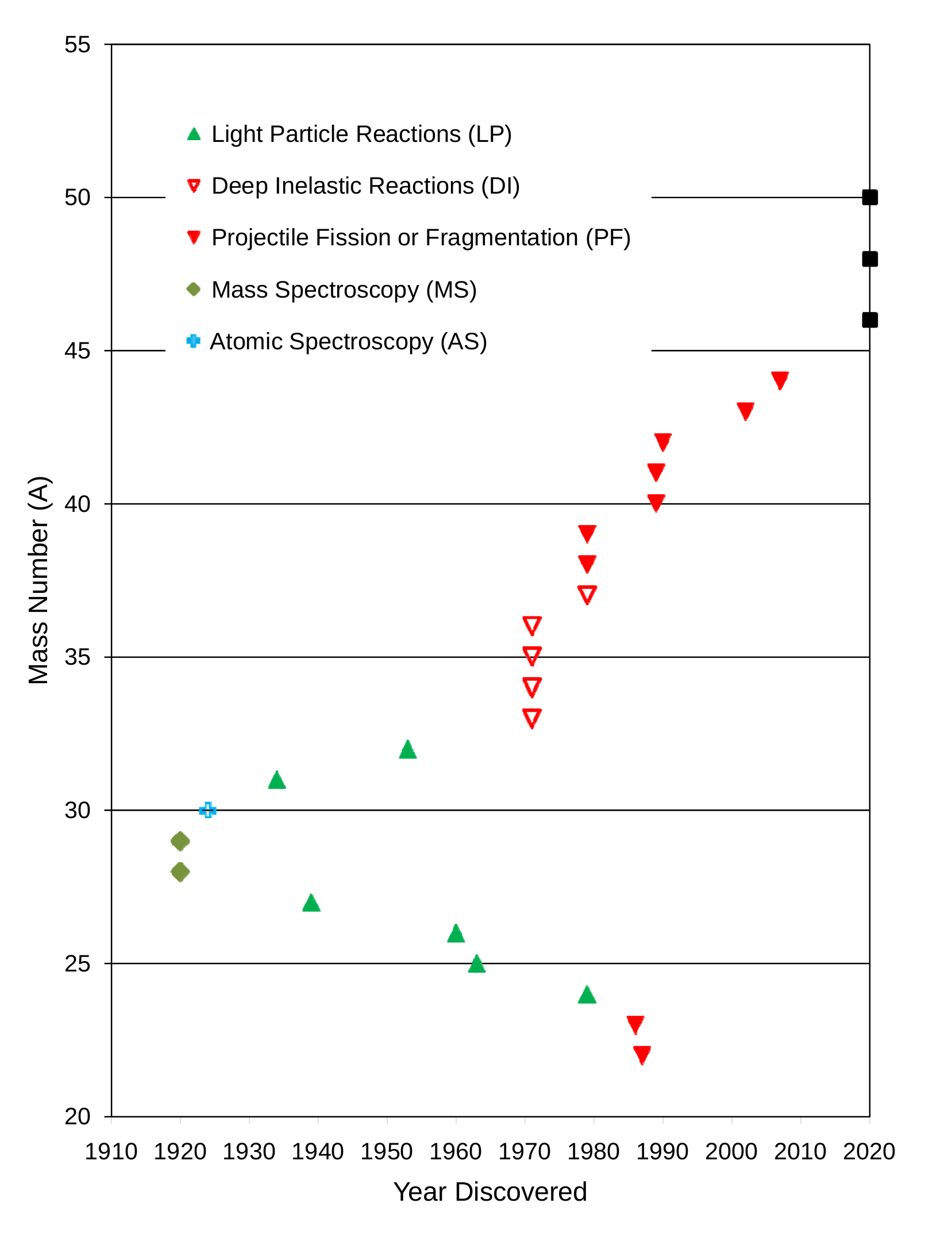}
	\caption{Silicon isotopes as a function of time when they were discovered. The different production methods are indicated. The solid black squares on the right hand side of the plot are isotopes predicted to be bound by the HFB-14 model.}
\label{f:year-si}
\end{figure}

Figure \ref{f:year-si} summarizes the year of first observation for all silicon isotopes identified by the method of discovery. The radioactive silicon isotopes were produced using light-particle reactions (LP), deep-inelastic reactions (DI), and projectile fragmentation of fission (PF). The stable isotopes were identified using mass spectroscopy (MS) and atomic spectroscopy (AS). Light particles are defined as incident particles with A$\le$4. The discovery of each silicon isotope is discussed in detail and a summary is presented in Table 1.

\subsubsection*{$^{22}$Si}

Saint-Laurent et al. discovered $^{22}$Si in 1987 in the paper ``Observation of a bound T$_z$=$-$3 nucleus: $^{22}$Si'' \cite{1987Sai01}.  A 85~MeV/u $^{36}$Ar beam was fragmented on a nickel target at GANIL and the projectile-like fragments were separated by the zero degree doubly achromatic LISE spectrometer. The isotopes were identified by measuring energy loss and time-of-flight. ``[The figure] shows the experimental results obtained after a 6-h run... The total number of $^{22}$Si observed in [the figure] is 161.''

\subsubsection*{$^{23}$Si}

$^{23}$Si was first reported in 1986 by Langevin et al. in ``Mapping of the proton drip-line up to Z = 20: Observation of the T$_z$=$-$5/2 series $^{23}$Si, $^{27}$S, $^{31}$Ar, and $^{35}$Ca'' \cite{1986Lan01}. A 77.4~MeV/u $^{40}$Ca beam was fragmented on a nickel target at GANIL and the projectile-like fragments were separated by the zero degree doubly achromatic LISE spectrometer. The isotopes were identified by measuring energy loss and time-of-flight. ``The bidimensional plot (see [the figure]) of $\sqrt{\Delta}$/t.o.f. (i.e. Z) versus t.o.f (i.e. A/Z) was inspected on-line to calibrate the particle identification... [The figure] shows the same bidimensional representation after 14 hours of integration time. The T$_z$ series $^{23}$Si, $^{27}$S, $^{31}$Ar, and $^{35}$Ca clearly becomes visible.''

\subsubsection*{$^{24}$Si}

In the 1979 paper ``Decay of a new isotope, $^{24}$Si: A test of the isobaric multiplet mass equation'' \"Ayst\"o et al. described the first observation of $^{24}$Si \cite{1979Ays01}. A 70 MeV $^{3}$He beam from the Berkeley 88-in. cyclotron bombarded magnesium targets and $^{24}$Si was produced in the reaction $^{24}$Mg($^3$He,3n). Beta-delayed protons from the recoil products were measured at the on-line mass separator system RAMA. ``The proton spectrum arising from the decay of $^{24}$Si after bombardment for 560 mC is shown in [the figure]. Only one peak is evident in the spectrum; it occurs at a laboratory energy of 3914$\pm$9~keV. Possible lower energy groups arising from positron decay to 1$^+$ states were not observed, a result partly due to the low detection efficiency of the telescope below 3 MeV. A half-life of 100$^{+90}_{-40}$~ms was estimated for the observed peak by comparing the $^{24}$Si focal plane yield to the yields of $^{20}$Na (t$_{1/2}$ = 446~ms), $^{24}$Al (2.07~s), $^{24}$Al$^m$ (129~ms) and $^{25}$Si (220~ms).'' This half-life agrees with the presently adopted value of 140(8)~ms.

\subsubsection*{$^{25}$Si}

In ``Observation of delayed proton radioactivity'' Barton et al. implied the observation of $^{25}$Si for the first time in 1963 \cite{1963Bar01}. The McGill Synchrocyclotron accelerated protons to 97 MeV which bombarded aluminum and SiO$_2$ targets. $^{25}$Si was identified by the observation of $\beta$-delayed protons in a silicon junction particle detector. ``In summary, it is concluded that all the observed peaks are due to delayed protons and that the spectra from Al and Si targets are due to the same radiations. It is concluded from activation and lifetime evidence that these protons are beta-delayed protons emitted from excited states of Al following the beta decay of Si$^{25}$.'' The half-life was subsequently measured by McPherson et al. \cite{1965McP01} who acknowledged the tentative observation by Barton et al.

\subsubsection*{$^{26}$Si}

$^{26}$Si was identified in 1960 by Robinson and Johnson in ``Decay of Si$^{26}$'' \cite{1960Rob01}. Magnesium targets were bombarded with an 8~MeV $^3$He beam from the Purdue 37-in cyclotron. $^{26}$Si was produced in the reaction $^{24}$Mg($^3$He,n) and identified by measuring decay curves and $\gamma$-ray spectra with a NaI(Tl) detector. ``An internally consistent argument based on the known decay characteristics of reaction products that may be expected from energy considerations, the results of half-life studies, experimental gamma spectra, and nuclear systematics can be made to support the conclusion that the (2.1$\pm$0.3)-sec halflife is that of Si$^{26}$ produced in the reaction Mg$^{24}$(He$^3$,n)Si$^{26}$, and a consistent decay scheme can be proposed.'' This half-life agrees with the presently accepted value of 2.234(13)~s. A previously measured 1.7~s half-life assigned to $^{26}$Si was not credited with the discovery because the identification was only suggested ``from a simple consideration of preferred reaction type[s] and estimates of threshold[s]'' \cite{1954Tyr01}.

\subsubsection*{$^{27}$Si}

Kuerti and Van Voorhis identified $^{27}$Si in 1939 as described in the paper ``Induced radioactivity produced by bombarding aluminum with protons'' \cite{1939Kue01}. Aluminum was bombarded with protons and $^{27}$Al was produced in the (p,n) charge exchange reaction. Excitation functions and activities were measured with an ionization chamber. Previously there had been some uncertainties about the assignment of a 6$-$7~min half-life to either $^{27}$Si or $^{29}$Al. ``However, all our attempts to find such a period have been completely unsuccessful, its intensity if present being at least ten thousand times weaker than would be predicted from (p,n) cross sections for neighboring elements. We have, however, found an activity of 3.7 seconds half-life which is produced in aluminum by protons.'' This half-life agrees with the presently accepted value of 4.16(2)~s. The previous measurements with the uncertain assignment were 7.5(15)~min \cite{1935Eck01}, 6.7(10)~min \cite{1935Fah01}, 6$-$7~min \cite{1936Ell01}, and 6.6(3)~min \cite{1937Mey01}. This half-life was later assigned to $^{29}$Al \cite{1939Bet01}. In addition, Curie and Joliet had assigned a 2.5~min half-life to $^{27}$Si \cite{1934Cur01,1934Jol01} which most likely was due to $^{28}$Al.

\subsubsection*{$^{28,29}$Si}

Aston discovered $^{28}$Si and $^{29}$Si in 1920 as reported in ``The constitution of the elements'' \cite{1920Ast02}. The isotopes were identified by measuring their mass spectra. ``The results obtained with silicon (atomic weight 28.3) are somewhat difficult to interpret, and lead to the conclusion that this element has isotopes 28 and 29, with possibly another 30.''

\subsubsection*{$^{30}$Si}

In the 1924 paper ``Isotope effects in the band spectra of boron monoxide and silicon nitride'' Mulliken reported the observation of $^{30}$Si \cite{1924Mul01}. Band spectra of silicon nitrate were measured. ``In agreement with theory for the heavier isotopes Si$^{29}$N and Si$^{30}$N, these weak heads lag behind the corresponding Si$^{28}$N heads more and more with
increasing distance toward the red from the central band. On the ultra-violet side of the central band, the isotope heads are concealed by the heavy shading of the Si$^{28}$N bands. Isotope 29 appears to be a little more abundant than isotope 30. There is no evidence of other isotopes in appreciable amounts.'' Previously Aston had only indicated the possibility of a stable $^{30}$Si isotope \cite{1920Ast02}.

\subsubsection*{$^{31}$Si}

$^{31}$Si was discovered by Fermi et al. in the 1934 article ``Artificial radioactivity produced by neutron bombardment'' \cite{1934Fer01}. Phosphorus targets were irradiated with neutrons from a 800 mCi radon beryllium source and activities were measured with Geiger-M\"uller counters following chemical separation. ``15$-$Phosphorus$-$This element shows a strong activity (i = 0.6) decaying with a period of about 3 hours... The 3 hours' active product could be chemically separated. For this purpose phosphorus was irradiated as a concentrated solution of phosphoric acid. This solution was afterwards diluted with water, adding sulphuric acid and a small amount of sodium silicate. The substance is dried up to render silica insoluble, and then dissolved in water and filtered. The activity is found with the silica. The nuclear reaction is then probably P$^{31}_{15}$ + n$^1_0$ = Si$^{31}_{13}$+ H$^1_1$.'' This half-life is consistent with the currently adopted value of 157.3(3)~min.

\subsubsection*{$^{32}$Si}

Lindner identified $^{32}$Si in the 1953 paper ``New nuclides produced in chlorine spallation'' \cite{1953Lin01}. A 340~MeV proton beam from the Berkeley 184-in. cyclotron bombarded sodium chloride targets. Beta absorption and decay curves were measured following chemical separation. ``The radiation properties of the beta-emitting nuclides Si$^{32}$ and Mg$^{28}$ are described. Si$^{32}$ was found to have a maximum probable half-life of 710 years, emitting beta-particles of E$_{max} \sim$ 100 kev... These data, therefore, establish the existence of the long-lived Si$^{32}$, which emits beta particles of about 100 kev and apparently no gamma radiation... It is unlikely that the half-life is as low as 100 years.'' The quoted range of possible half-lives includes the currently adopted value of 132(13)~y. A previous search for $^{32}$Si was unsuccessful \cite{1953Lin02} while Turkevich and Tompkins set an upper limit for the abundance of $^{32}$Si in natural silicon \cite{1953Tur01}.

\subsubsection*{$^{33-36}$Si}

Artukh et al. discovered $^{33}$Si, $^{34}$Si, $^{35}$Si, and $^{36}$Si in the 1971 paper ``New isotopes $^{29,30}$Mg, $^{31,32,33}$Al, $^{33,34,35,36}$Si, $^{35,36,37,38}$P, $^{39,40}$S, and $^{41,42}$Cl produced in bombardment of a $^{232}$Th target with 290 MeV $^{40}$Ar ions'' \cite{1971Art01}. A 290 MeV $^{40}$Ar beam from the Dubna 310 cm heavy-ion cyclotron bombarded a metallic $^{232}$Th. Reaction products were separated and identified with a magnetic spectrometer and a surface barrier silicon telescope. ``Apart from the nucleides already known, 17 new nucleides, namely: $^{29,30}$Mg, $^{31,32,33}$Al, $^{33,34,35,36}$Si, $^{35,36,37, 38}$P, $^{39,40}$S and $^{41,42}$Cl have been reliably detected.''

\subsubsection*{$^{37}$Si}

In 1979 $^{37}$Si was discovered by  Auger et al. in ``Observation of new nuclides $^{37}$Si, $^{40}$P, $^{41}$S, $^{42}$S produced in deeply inelastic reactions induced by $^{40}$Ar on $^{238}$U'' \cite{1979Aug01}.  A 263 MeV $^{40}$Ar beam from the Orsay ALICE facility bombarded a UF$_4$ target and reaction products were measured with a triple silicon solid state counter telescope. ``Four new neutron-rich nuclides, $^{37}$Si, $^{40}$P, $^{41-42}$S have been observed as a result of deep inelastic collisions. The nuclide identification combined two independent time of flight measurements as well as two ($\Delta$E $\times$ E) informations and was quite unambiguous.''

\subsubsection*{$^{38,39}$Si}

The first observation of $^{38}$Si and $^{39}$Si was reported by Westfall et al. in ``Production of neutron-rich nuclides by fragmentation of 212-MeV/amu $^{48}$Ca'' in 1979 \cite{1979Wes01}. $^{48}$Ca ions (212 MeV/nucleon) from the Berkeley Bevalac were fragmented on a beryllium target. The fragments were selected by a zero degree spectrometer and identified in a telescope consisting of 12 Si(Li) detectors, 2 position-sensitive Si(Li) detectors, and a veto scintillator. ``In this letter, we present the first experimental evidence for the particle stability of fourteen nuclides $^{22}$N, $^{26}$F, $^{33,34}$Mg, $^{36,37}$Al, $^{38,39}$Si, $^{41,42}$P, $^{43,44}$S, and $^{44,45}$Cl produced in the fragmentation of 212-MeV/amu $^{48}$Ca.''

\subsubsection*{$^{40,41}$Si}

Guillemaud-Mueller et al. announced the discovery of $^{40}$Si and $^{41}$Si in the 1989 article ``Observation of new neutron rich nuclei $^{29}$F, $^{35,36}$Mg, $^{38,39}$Al, $^{40,41}$Si, $^{43,44}$P, $^{45-47}$S, $^{46-49}$Cl, and $^{49-51}$Ar from the interaction of 55 MeV/u $^{48}$Ca+Ta'' \cite{1989Gui01}. A 55~MeV/u $^{48}$Ca beam was fragmented on a tantalum target at GANIL and the projectile-like fragments were separated by the zero degree doubly achromatic LISE spetrometer. ``[The figure] represents the two-dimensional plot (energy loss versus time-of-flight) obtained under these conditions after 40~h integration time with an average intensity of 150 enA. The new species $^{35,36}$Mg, $^{38,39}$Al, $^{40,41}$Si, $^{43,44}$P, $^{45,46,47}$S, $^{46,47,48,49}$Cl, and $^{49,50,51}$Ar are clearly visible.''

\subsubsection*{$^{42}$Si}

Lewitowicz et al. discovered $^{42}$Si in the 1990 paper ``First observation of the neutron-rich nuclei $^{42}$Si, $^{45,46}$P, $^{48}$S, and $^{51}$Cl from the interaction of 44 MeV/u $^{48}$Ca + $^{64}$Ni'' \cite{1990Lew01}. A 44~MeV/u $^{48}$Ca beam was fragmented on a $^{64}$Ni  target at GANIL and the projectile-like fragments were separated by the zero degree doubly achromatic LISE spectrometer. ``The isotopes of $^{42}$Si, $^{45,46}$P, $^{48}$S, and $^{51}$Cl are identified for the first time.''

\subsubsection*{$^{43}$Si}

In the 2002 article ``New neutron-rich isotopes, $^{34}$Ne, $^{37}$Na and $^{43}$Si, produced by fragmentation of a 64A MeV $^{48}$Ca beam'' Notani et al. described the first observation of $^{43}$Si \cite{2002Not01}. The RIKEN ring cyclotron accelerated a $^{48}$Ca beam to 64 MeV/nucleon which was then fragmented on a tantalum target. The projectile fragments were analyzed with the RIPS spectrometer. ``[Part (a) of the figure] shows a two-dimensional plot of A/Z versus Z, obtained from the data accumulated with the $^{40}$Mg B$\rho$ setting, while [part (b)] is for the $^{43}$Si setting. The integrated beam intensities for the two settings are 6.9$\times$10$^{16}$ and 1.7$\times$10$^{15}$ particles, respectively. The numbers of events observed for three new isotopes, $^{34}$Ne, $^{37}$Na and $^{43}$Si, were 2, 3 and 4, respectively.''

\subsubsection*{$^{44}$Si}

Tarasov et al. reported the first observation of $^{44}$Si in ``New isotope $^{44}$Si and systematics of the production cross sections of the most neutron-rich nuclei'' in 2007 \cite{2007Tar01}. A $^9$Be target was bombarded with a 142 MeV/nucleon $^{48}$Ca from the NSCL coupled cyclotron facility. $^{44}$Si was identified with the A1900 fragment separator. ``The study of the production of the most neutron-rich silicon isotopes provided evidence for the existence of a new isotope, $^{44}$Si, in a high energy reaction that requires the net transfer of two neutrons to the projectile.''

\subsection{Phosphorus}\vspace{0.0cm}

The observation of 21 phosphorus isotopes has been reported so far, including 1 stable, 5 proton-rich, and 15 neutron-rich isotopes. The proton dripline has been reached because as stated in \cite{2004Tho01} the particle identification plot in the 1986 paper by Langevin et al. clearly showed that $^{25}$P is beyond the dripline and has a lifetime shorter than the time-of-flight of 170~ns. According to the HFB-14 model \cite{2007Gor01}, four more phosphorus neutron-rich isotopes could be bound ($^{47-49}$P and $^{51}$P).

\begin{figure}
	\centering
	\includegraphics[scale=0.7]{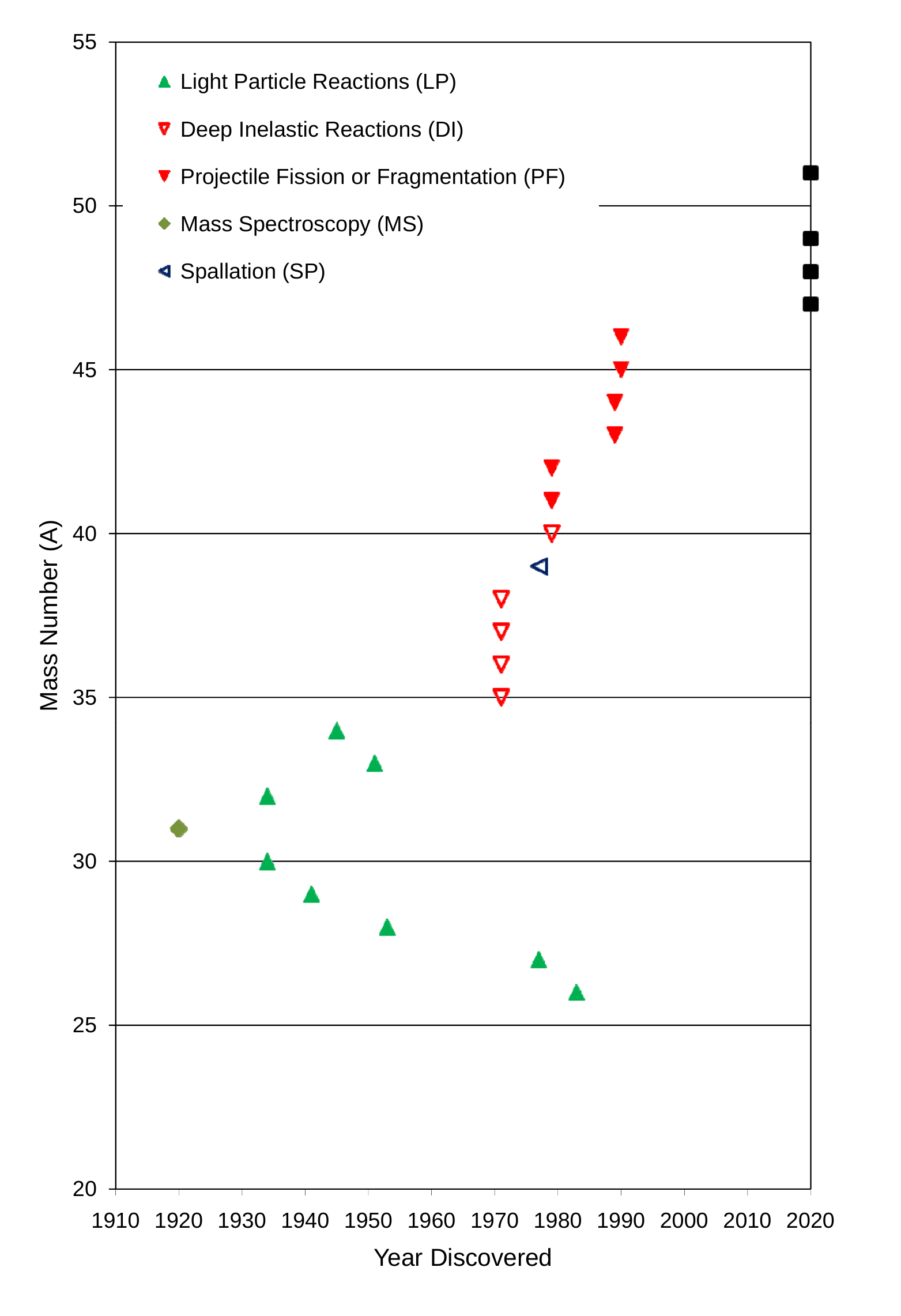}
	\caption{Phosphorus isotopes as a function of time when they were discovered. The different production methods are indicated. The solid black squares on the right hand side of the plot are isotopes predicted to be bound by the HFB-14 model.}
\label{f:year-p}
\end{figure}

Figure \ref{f:year-p} summarizes the year of first observation for all phosphorus isotopes identified by the method of discovery. The radioactive phosphorus isotopes were produced using light-particle reactions (LP), deep-inelastic reactions (DI), spallation (SP), and projectile fragmentation of fission (PF). The stable isotopes were identified using mass spectroscopy (MS). Light particles are defined as incident particles with A$\le$4. The discovery of each phosphorus isotope is discussed in detail and a summary is presented in Table 1.

\subsubsection*{$^{26}$P}

The discovery of $^{26}$P was reported in 1983 by Cable et al. in ``Beta-delayed proton decay of the T$_z$=$-$2 isotope $^{26}$P'' \cite{1983Cab01}. $^{28}$Si was bombarded with a 110$-$130~MeV $^3$He beam from the Berkeley 88-in. cyclotron and $^{26}$P was produced in the reaction $^{28}$Si($^3$He,p4n). Recoil products were collected on a rotating wheel and $\beta$-delayed protons were measured in a three-element semiconductor telescope. ``A rough half-life of 20$^{+35}_{-15}$~ms was determined for the 7.269~MeV proton group by varying the wheel rotational speed and observing the relative yields of
the various activities present. This method could, in principle, yield a precise half-life measurement but does not do so here due to the low yield of $^{26}$P. As shown in [the figure], the 7.269$\pm$0.015~MeV group can be attributed to the isospin forbidden decay of the lowest T = 2 state in $^{26}$Si (fed by the superallowed $\beta^+$-decay of the T = 2 ground state of $^{26}$P) to the ground state of $^{25}$Al.'' This half-life is consistent with the presently adopted value of 30(25)~ms.

\subsubsection*{$^{27}$P}

Benenson et al. observed $^{27}$P in 1977 as reported in ``Mass of $^{27}$P and $^{31}$Cl'' \cite{1977Ben01}. A $^{32}$S target was bombarded with a 70~MeV $^3$He beam and $^{27}$P was formed in a ($^3$He,$^8$Li) reaction. The ejectiles were measured with a double proportional counter and a thin plastic scintillator at the focal plane of a split-pole spectrograph. ``The spectrum from the $^{32}$S($^3$He,$^8$Li)$^{27}$P reaction at 7$^\circ$ is given in [the figure]. The $^{27}$P ground state is clearly evident although not at all strong.''

\subsubsection*{$^{28}$P}

In 1953 $^{28}$P was reported by Glass et al. in ``The short-lived radioisotopes P$^{28}$ and Cl$^{32}$'' \cite{1953Gla01}. Protons were accelerated to 20 MeV by the UCLA cyclotron and bombarded silicon targets. $^{28}$P was produced in (p,n) charge exchange reactions and identified by measuring $\gamma$-rays with a NaI crystal. ``The half-life of the P$^{28}$ was found to be 0.280$\pm$0.010 second. It emits positrons and gamma-radiation up to an energy of 7 Mev.'' This half-life is in agreement with the currently adopted value of 270.3(5)~ms.

\subsubsection*{$^{29}$P}

White et al. described the observation of $^{29}$P in 1941 in the paper ``Positrons from light nuclei'' \cite{1941Whi01}. Protons from the Princeton cyclotron bombarded silicon targets. Beta-rays were measured in a cloud chamber and the half-life was recorded by taking photographs of a stop watch dial and the image of the fiber of a projection-type Lauritsen electroscope. ``A large number of targets were used in rotation so that the P$^{30}$ activity would not build up after repeated short exposures to the beam. It was hoped that the energy of P$^{29}$ would be sufficiently higher than that of P$^{30}$ so it could be distinguished. That this was possible may be seen from [the figure], where the momentum spectrum of all positrons from the two reactions Si$^{29,30}$(p,n)P$^{29,30}$ is plotted as well as the upper end of the spectrum of positrons from P$^{30}$ alone, the latter being obtained when the proton energy was below the threshold for production of P$^{29}$.'' The reported half-life of 4.6(2)~s agrees with the presently adopted value of 4.142(15)~s.

\subsubsection*{$^{30}$P}

Curie and Joliot presented first evidence of $^{30}$P in 1934 in ``Un nouveau type de radioactivit\'e'' \cite{1934Cur01}. Aluminum, boron, and magnesium samples were irradiated by polonium $\alpha$-particles and their activities were measured with a Geiger M\"uller counter as a function of time. ``Nous pla\c{c}ons une feuille d'aluminium \`a 1$^{mm}$ d'une source de polonium. L'aluminium ayant \'et\'e irradi\'e pendant 10 minutes environ, nous le pla\c{c}ons au-dessus d'un compteur de Geiger M\"uller portant un orifice ferm\'e par un \'ecran de 7/100$^e$ millim\'etre d'aluminium. Nous observons que la feuille \'emet un rayonnement dont l'intensit\'e d\'ecro\^it exponentiellement en fonction du temps avec une p\'eriode de 3 minutes 15 secondes... Ces exp\'eriences montrent l'exitence d'un nouveau type de radioactivit\'e avec \'emission d'\'electrons positifs. Nous pensons que le processus d'\'emission serait le suivant pour l'aluminium: $^{27}_{13}$Al + $^4_2$He = $^{30}_{15}$P + $^1_0$n.'' [We place a 1~mm aluminum sheet in front of a polonium source. After the aluminum was irradiated for 10 minutes, we place it on top of a Geiger M\"uller counter and an aluminum screen with a 7/100~mm aperture. We observe that the sheet emits radiation whose intensity decreases exponentially with a period of 3 minutes and 15 seconds... These experiments show the existence of a new type of radioactivity with the emission of positive electrons. We propose the following emission process: $^{27}_{13}$Al + $^4_2$He = $^{30}_{15}$P + $^1_0$n.] This half-life is close to the currently adopted value of 2.498(4)~min.

\subsubsection*{$^{31}$P}

Aston discovered $^{31}$P in 1920 as reported in ``The constitution of the elements'' \cite{1920Ast02}. The isotopes were identified by measuring their mass spectra. ``Phosphorus (atomic weight 31.04) and arsenic (atomic weight 74.96) are also apparently simple elements of masses 31 and 75 respectively.''

\subsubsection*{$^{32}$P}

$^{32}$P was discovered by Fermi et al. in the 1934 article ``Artificial radioactivity produced by neutron bombardment'' \cite{1934Fer01}. Phosphorus targets were irradiated with neutrons from a 800 mCi radon beryllium source and activities were measured with Geiger-M\"uller counters following chemical separation. ``15$-$Sulphur: Sulphur shows a fairly strong activity, decaying with a period of about 13 days (rather inaccurately measured). Half-value absorption thickness of the $\beta$-rays 0.10~gm/cm$^2$. A chemical separation of the active product was carried out as follows: irradiated sulphuric acid was diluted, a trace of sodium phosphate added, and phosphorus precipitated as phosphomolibdate by addition of ammonium molibdate. The activity was found in the precipitate. We think, in consequence, that the nuclear reaction is S$^{32}_{16}$ + n$^1_0$ = P$^{32}_{15}$+ H$^1_1$.'' This half-life agrees with the currently adopted value of 14.263(3)~d.

\subsubsection*{$^{33}$P}

Sheline et al. reported the observation of $^{33}$P in the 1951 paper ``The nuclide P$^{33}$ and the P$^{32}$ spectrum'' \cite{1951She01}. The Chicago betatron was used to produce 48~MeV $\gamma$-rays which bombarded sulfur, sodium sulfide and lithium chloride targets. $^{33}$P was produced in the photo-nuclear reactions $^{34}$S($\gamma$,p), $^{35}$Cl($\gamma$,2p), and $^{37}$Cl($\gamma$,$\alpha$). Beta-rays were recorded with an end window Geiger tube and absorption curves were measured. ``The P$^{33}$ activity has a 25$\pm$2 day half-life with a 0.27$\pm$0.02-Mev negative beta-ray. There is less than one (0.5-Mev) gamma for every fifteen betas.'' This half-life agrees with the presently adopted value of 25.34(12)~d. A previously measured 22(5)~s half-life was reported as a meeting abstract \cite{1951Yaf01} and was evidently incorrect.

\subsubsection*{$^{34}$P}

In 1945 $^{34}$P was identified by Z\"unti and Bleuler as described in ``\"Uber zwei Aktivit\"aten S$^{37}$ und P$^{34}$, die durch schnelle Neutronen in Chlor induziert werden'' \cite{1945Zun01}. Fast neutrons produced by a tensator were used to bombard chlorine targets. Beta- and gamma-ray spectra were measured following chemical separation. ``Da der n,$\alpha$-Prozess bei Cl$^{35}$ auf den bekannten langlebigen P$^{32}$ f\"uhrt, kann es sich nur um die Reaktion Cl$^{37}$(n,$\alpha$)P$^{34}$ handeln.'' [Because the n,$\alpha$-process on Cl$^{35}$ leads to the known longlived P$^{32}$, it can only be due to the reaction Cl$^{37}$(n,$\alpha$)P$^{34}$.] The measured half-life of 12.4(2)~s agrees with the currently adopted value of 12.43(8)~s. Previously a 14.7~s half-life was reported to result from either $^{34}$P or $^{37}$S \cite{1942Hub01}, and a 12.7~s half-life was assigned to a phosphorus isotope with mass larger than 31 \cite{1940Cor01}.

\subsubsection*{$^{35-38}$P}

Artukh et al. discovered $^{35}$P, $^{36}$P, $^{37}$P, and $^{38}$P in the 1971 paper ``New isotopes $^{29,30}$Mg, $^{31,32,33}$Al, $^{33,34,35,36}$Si, $^{35,36,37,38}$P, $^{39,40}$S, and $^{41,42}$Cl produced in bombardment of a $^{232}$Th target with 290 MeV $^{40}$Ar ions'' \cite{1971Art01}. A 290 MeV $^{40}$Ar beam from the Dubna 310 cm heavy-ion cyclotron bombarded a metallic $^{232}$Th. Reaction products were separated and identified with a magnetic spectrometer and a surface barrier silicon telescope. ``Apart from the nucleides already known, 17 new nucleides, namely: $^{29,30}$Mg, $^{31,32,33}$Al, $^{33,34,35,36}$Si, $^{35,36,37, 38}$P, $^{39,40}$S and $^{41,42}$Cl have been reliably detected.'' Less than two months later Grimm and Herzog independently reported the first half-life of 45(2)~s for $^{35}$P \cite{1971Gri01}.

\subsubsection*{$^{39}$P}

$^{39}$P was discovered by Butler et al. in ``Observation of the new nuclides $^{27}$Ne, $^{31}$Mg, $^{32}$Mg, $^{34}$Al, and $^{39}$P'' in 1977 \cite{1977But01}. $^{39}$P was produced in the spallation reaction of 800 MeV protons from the Clinton P. Anderson Meson Physics Facility LAMPF on a uranium target. The spallation fragments were identified with a silicon $\Delta$E-E telescope and by time-of-flight measurements. ``All of the stable and known neutron-rich nuclides (except $^{24}$O and the more neutron-rich Na isotopes) are seen. The five previously unobserved neutron-rich nuclides $^{27}$Ne, $^{31}$Mg, $^{32}$Mg, $^{34}$Al, and $^{39}$P are clearly evident. Each of these peaks contains ten or more events.''

\subsubsection*{$^{40}$P}

In 1979 $^{40}$P was discovered by  Auger et al. in ``Observation of new nuclides $^{37}$Si, $^{40}$P, $^{41}$S, $^{42}$S produced in deeply inelastic reactions induced by $^{40}$Ar on $^{238}$U'' \cite{1979Aug01}.  A 263 MeV $^{40}$Ar beam from the Orsay ALICE facility bombarded a UF$_4$ target and reaction products were measured with a triple silicon solid state counter telescope. ``Four new neutron-rich nuclides, $^{37}$Si, $^{40}$P, $^{41-42}$S have been observed as a result of deep inelastic collisions. The nuclide identification combined two independent time of flight measurements as well as two ($\Delta$E $\times$ E) informations and was quite unambiguous.''

\subsubsection*{$^{41,42}$P}

The first observation of $^{41}$P and $^{42}$P was reported by Westfall et al. in ``Production of neutron-rich nuclides by fragmentation of 212-MeV/amu $^{48}$Ca'' in 1979 \cite{1979Wes01}. $^{48}$Ca ions (212 MeV/nucleon) from the Berkeley Bevalac were fragmented on a beryllium target. The fragments were selected by a zero degree spectrometer and identified in a telescope consisting of 12 Si(Li) detectors, 2 position-sensitive Si(Li) detectors, and a veto scintillator. ``In this letter, we present the first experimental evidence for the particle stability of fourteen nuclides $^{22}$N, $^{26}$F, $^{33,34}$Mg, $^{36,37}$Al, $^{38,39}$Si, $^{41,42}$P, $^{43,44}$S, and $^{44,45}$Cl produced in the fragmentation of 212-MeV/amu $^{48}$Ca.''

\subsubsection*{$^{43,44}$P}

Guillemaud-Mueller et al. announced the discovery of $^{43}$P and $^{44}$P in the 1989 article ``Observation of new neutron rich nuclei $^{29}$F, $^{35,36}$Mg, $^{38,39}$Al, $^{40,41}$Si, $^{43,44}$P, $^{45-47}$S, $^{46-49}$Cl, and $^{49-51}$Ar from the interaction of 55 MeV/u $^{48}$Ca+Ta'' \cite{1989Gui01}. A 55~MeV/u $^{48}$Ca beam was fragmented on a tantalum target at GANIL and the projectile-like fragments were separated by the zero degree doubly achromatic LISE spetrometer. ``[The figure] represents the two-dimensional plot (energy loss versus time-of-flight) obtained under these conditions after 40~h integration time with an average intensity of 150 enA. The new species $^{35,36}$Mg, $^{38,39}$Al, $^{40,41}$Si, $^{43,44}$P, $^{45,46,47}$S, $^{46,47,48,49}$Cl, and $^{49,50,51}$Ar are clearly visible.''

\subsubsection*{$^{45,46}$P}

Lewitowicz et al. discovered $^{45}$P and $^{46}$P in the 1990 paper ``First observation of the neutron-rich nuclei $^{42}$Si, $^{45,46}$P, $^{48}$S, and $^{51}$Cl from the interaction of 44 MeV/u $^{48}$Ca + $^{64}$Ni'' \cite{1990Lew01}. A 44~MeV/u $^{48}$Ca beam was fragmented on a $^{64}$Ni  target at GANIL and the projectile-like fragments were separated by the zero degree doubly achromatic LISE spectrometer. ``The isotopes of $^{42}$Si, $^{45,46}$P, $^{48}$S, and $^{51}$Cl are identified for the first time.''

\subsection{Sulfur}\vspace{0.0cm}

The observation of 22 sulfur isotopes has been reported so far, including 4 stable, 5 proton-rich, and 13 neutron-rich isotopes. No specific searches for the existence of $^{26}$S have been reported and so it could potentially still be observed \cite{2004Tho01}. The latest mass evaluation predicts the one-proton separation energy of $^{26}$S to be 190(360)~keV. According to the HFB-14 model \cite{2007Gor01}, four more sulfur neutron-rich isotopes could be bound ($^{49,50}$S, $^{52}$S, and $^{54}$S).

\begin{figure}
	\centering
	\includegraphics[scale=0.7]{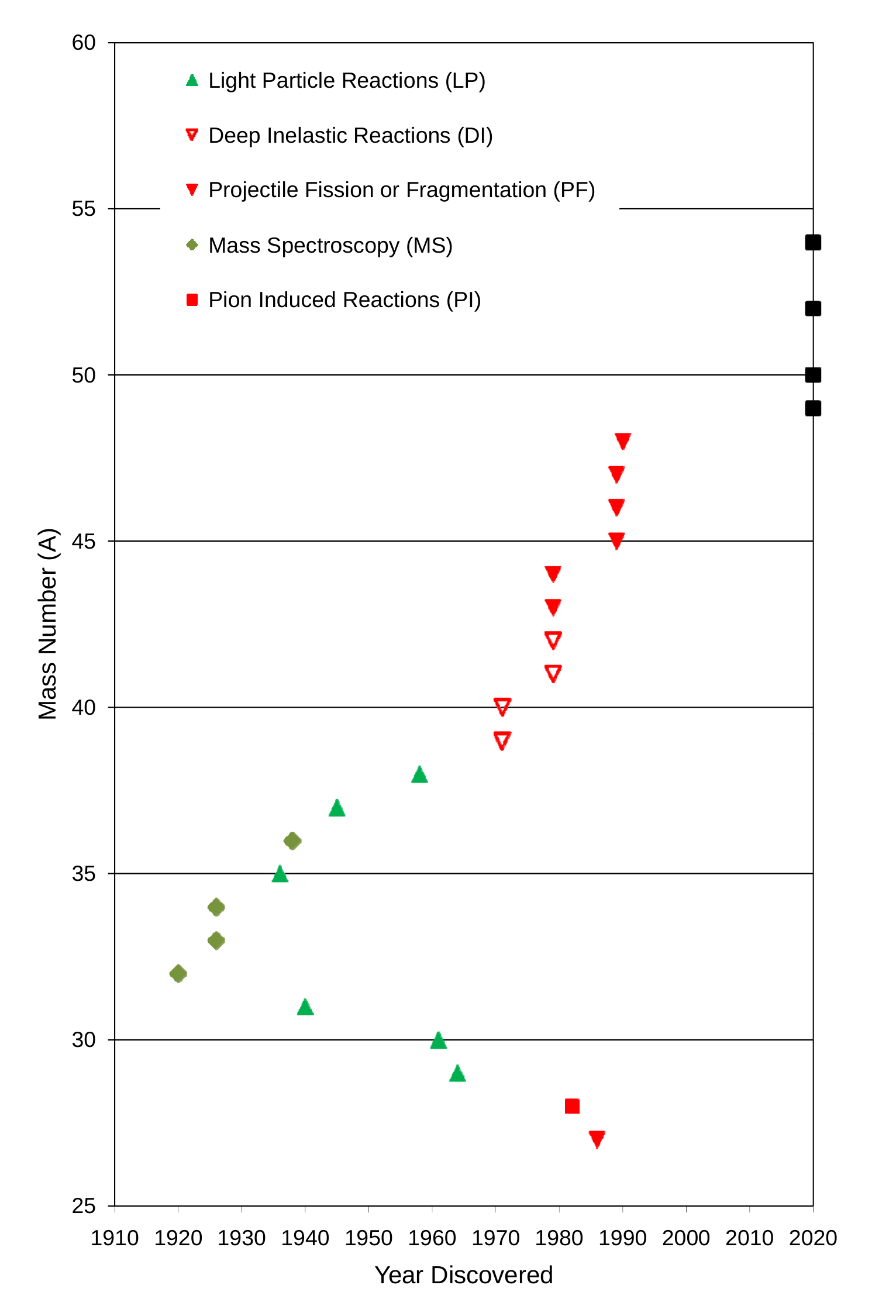}
	\caption{Sulfur isotopes as a function of time when they were discovered. The different production methods are indicated. The solid black squares on the right hand side of the plot are isotopes predicted to be bound by the HFB-14 model.}
\label{f:year-s}
\end{figure}

Figure \ref{f:year-s} summarizes the year of first observation for all sulfur isotopes identified by the method of discovery. The radioactive sulfur isotopes were produced using light-particle reactions (LP), pion-induced reactions (PI), spallation (SP), and projectile fragmentation of fission (PF). The stable isotopes were identified using mass spectroscopy (MS). Light particles are defined as incident particles with A$\le$4. The discovery of each sulfur isotope is discussed in detail and a summary is presented in Table 1.

\subsubsection*{$^{27}$S}

$^{27}$S was first reported in 1986 by Langevin et al. in ``Mapping of the proton drip-line up to Z = 20: Observation of the T$_z$=$-$5/2 series $^{23}$Si, $^{27}$S, $^{31}$Ar, and $^{35}$Ca'' \cite{1986Lan01}. A 77.4~MeV/u $^{40}$Ca beam was fragmented on a nickel target at GANIL and the projectile-like fragments were separated by the zero degree doubly achromatic LISE spectrometer. The isotopes were identified by measuring energy loss and time-of-flight. ``The bidimensional plot (see [the figure]) of $\sqrt{\Delta}$/t.o.f. (i.e. Z) versus t.o.f (i.e. A/Z) was inspected on-line to calibrate the particle identification... [The figure] shows the same bidimensional representation after 14 hours of integration time. The T$_z$ series $^{23}$Si, $^{27}$S, $^{31}$Ar, and $^{35}$Ca clearly becomes visible.''

\subsubsection*{$^{28}$S}

Morris et al. discovered $^{28}$S in ``Target mass dependence of isotensor double charge exchange: Evidence for deltas in nuclei'' in 1982 \cite{1982Mor01}. $^{28}$S was produced by the pion induced double charge exchange reaction $^{28}$Si($\pi ^+$,$\pi ^-$) and the negative pions were analyzed with the Energetic Pion Channel and Spectrometer EPICS. ``Byproducts of the present measurements are values of the masses of $^{28}$S and $^{40}$Ti. Our measured mass excesses are 4.13$\pm$0.16 and $-$8.79$\pm$0.16 MeV for $^{28}$S and $^{40}$Ti, respectively.''

\subsubsection*{$^{29}$S}

In 1964 Hardy and Verrall reported the first observation of $^{29}$S in ``Delayed protons following the decay of S$^{29}$'' \cite{1964Har02}. A thin sulfur target was inserted on a radial probe into the circulating proton beam of the McGill synchrocyclotron. Beta delayed protons were measured with a surface barrier silicon detector. ``Typical decay curves for the three main peaks are shown in [the figure]. The data for the 5.59~MeV peak has been corrected for the small Si$^{25}$ peak (5.62~MeV). From such curves, the half-life we adopt for S$^{29}$ is 195$\pm$8~msec.'' This half-life agrees with the presently adopted value of 187(4)~ms.

\subsubsection*{$^{30}$S}

Robinson et al. discovered $^{30}$S as described in the 1961 article ``Decay of a new isotope, S$^{30}$'' \cite{1961Rob02}. Natural silicon targets were irradiated with an 8~MeV $^3$He beam from the Purdue cyclotron. $^{30}$S was formed in a ($^3$He,n) reaction and identified by measuring $\beta$- and $\gamma$-ray spectra. ``A radioisotope with a (1.35$\pm$0.10)-sec half-life is produced in the bombardment of high-purity silicon with 8-Mev He$^3$ ions. The observed half-life is that of the new isotope S$^{30}$ produced in the reaction Si$^{28}$(He$^3$,n)S$^{30}$.'' This half-life is close to the currently adopted value of 1.178(5)~s.

\subsubsection*{$^{31}$S}

King and Elliott identified $^{31}$S in ``Short-lived radioactivities of $_{14}$Si$^{27}$, $_{16}$S$^{31}$, and $_{18}$A$^{35}$'' in 1940 \cite{1940Kin01}. Magnesium targets were bombarded with 16~MeV $\alpha$ particles and the resulting activities were measured with a multiple Geiger counter circuit. ``In an attempt to extend the well-known series of radioactive elements characterized by the formula Z $-$ N = 1, the following new reactions have been observed:... Reaction: $_{14}$Si$^{28}$($\alpha$,n)$_{16}$S$^{31}$; Half-life: 3.18~s.'' This half-life is near the presently accepted value of 2.572(13)~s.

\subsubsection*{$^{32}$S}

Aston discovered $^{32}$S in 1920 as reported in ``The constitution of the elements'' \cite{1920Ast02}. The isotopes were identified by measuring their mass spectra. ``Sulphur (atomic weight 32.06) has a predominant constituent 32. Owing to possible hydrogen compounds the data are as yet insufficient to give a decision as to the presence of small quantities of isotopes of higher mass suggested by the atomic weight.''

\subsubsection*{$^{33,34}$S}

In the 1926 article ``The isotopes of sulphur'' Aston reported the discovery of $^{33}$S and $^{34}$S \cite{1926Ast01}. The discovery was possible due to an improved resolving power of the new Cavendish mass spectrograph. Previously, Aston was not able to identify sulfur isotopes other than $^{32}$S due to possible hydrogen compounds \cite{1920Ast02}. ``The matter has now been put beyond reasonable doubt by the negative mass-spectrum obtained by using pure SO$_2$ and exposing for an hour with both fields reversed. All three lines were visible, and again showed the same intensity relations. Sulphur is therefore a triple element like the two even ones, magnesium and silicon, which precede it in the periodic table. The lightest mass-number is for the most abundant in all three cases. S$^{34}$ appears to be about three times as abundant as S$^{33}$; the two together probably amount to about 3 per cent. of the whole.''

\subsubsection*{$^{35}$S}

Anderson discovered $^{35}$S in 1936 as described in ``Ein radioaktives Isotop des Schwefels'' \cite{1936And01}. Carbon tetrachloride was irradiated with neutrons from a radium emanation ($^{222}$Rn) $-$ beryllium source. The resulting activity was measured following chemical separation. ``Diese drei Proben waren alle aktiv und ergaben die gleiche Halbwertszeit von 80 Tagen mit einer gesch\"atzten Unsicherheit von $\pm$10 Tagen. Es wurde nur eine Halbwertszeit beobachtet. Wie oben erw\"ahnt, darf man vermuten, dass dies dem Isotop $_{16}$S$^{35}$ eigen ist.'' [These three probes were all active and had the same half-life of 80 days with an estimated uncertainty of $\pm$10~days. Only one half-life was observed. As mentioned above, one can assume it corresponds to the half-life of the isotope $_{16}$S$^{35}$.] This half-life agrees with the currently adopted value of 87.51(12)~d.

\subsubsection*{$^{36}$S}

Nier reported the discovery of $^{36}$S in 1938 in his paper  ``The isotopic constitution of calcium, titanium, sulfur and argon'' \cite{1938Nie01}. SO$_2$ flowed into the tube of a mass spectrometer and positive ion peaks of SO$_2^+$, SO$^+$ and S$^+$ were used to identify $^{36}$S. ``A new sulphur isotope, S$^{36}$, was discovered, having an abundance 1/6,000 that of S$^{32}$.''

\subsubsection*{$^{37}$S}

In 1945, $^{37}$S was identified by Z\"unti and Bleuler as described in ``\"Uber zwei Aktivit\"aten S$^{37}$ und P$^{34}$, die durch schnelle Neutronen in Chlor induziert werden'' \cite{1945Zun01}. Fast neutrons produced by a tensator were used to bombard chlorine targets. Beta- and gamma-ray spectra were measured following chemical separation. ``Bei den Messungen an diesem Phosphorisotop bemerkten wir die Anwesenheit einer l\"angern Periode. Subtrahiert man von der Abklingkurve des bestrahlten Chlors die bekannten Aktivit\"aten von P$^{34}$, Cl$^{34}$, Cl$^{35}$, P$^{32}$ und S$^{35}$, so bleibt ein rein exponentieller Abfall mit 5,0 min Halbwertszeit \"ubrig. Die chemische Abtrennung zeigt, dass diese Aktivit\"at einem Schwefelisotop zukommt und zwar dem S$^{37}$, da nach Kamen der S$^{35}$ mit einer 88-Tage-Periode zerf\"allt.'' [During the measurements of this phosphor isotope we noticed the presence of a longer period. After subtraction of the known activities of P$^{34}$, Cl$^{34}$, Cl$^{35}$, P$^{32}$ and S$^{35}$ from the decay curve of the irradiated chlorine a pure exponential decay with a half-life of 5.0~min remains. The chemical separation shows, that this activity is due to a sulfur isotope, specifically S$^{37}$, because Kamen had shown that S$^{35}$ decays with a period of 88 days.] The measured half-life of 5.04(2)~min agrees with the currently adopted value of 5.05(2)~min. Previously a 14.7~s half-life was reported to result from either $^{34}$P or $^{37}$S \cite{1942Hub01}.

\subsubsection*{$^{38}$S}

$^{38}$S was identified in 1958 by Nethaway and Caretto in ``New isotope, sulfur-38'' \cite{1958Net01}. A 48-MeV $\alpha$-particle beam from the Berkeley 60-in cyclotron bombarded reagent-grade NaCl crystals and $^{38}$S was formed in the reaction $^{37}$Cl($\alpha$,3p). Beta-activity was measured with a proportional counter following chemical separation. ``All the chlorine samples were observed to decay with a single 37-minute half-life. The observed counting rates were extrapolated to the time of the sulfur-chlorine separation and then corrected for the chemical yield of the chlorine carrier added and for the loss of sulfur in each separation step. In [the figure] these results are presented in a plot of corrected chlorine activity versus the time of separation. The slope of the line indicates that the parent of Cl$^{38}$ has a half-life of about 172 minutes, and the repeated chemical isolation of Cl$^{38}$ from the sulfur fraction verifies the assignment as S$^{38}$.'' This half-life agrees with the presently adopted value of 170.3(7)~min.

\subsubsection*{$^{39,40}$S}

Artukh et al. discovered $^{39}$S and $^{40}$S in the 1971 paper ``New isotopes $^{29,30}$Mg, $^{31,32,33}$Al, $^{33,34,35,36}$Si, $^{35,36,37,38}$P, $^{39,40}$S, and $^{41,42}$Cl produced in bombardment of a $^{232}$Th target with 290 MeV $^{40}$Ar ions'' \cite{1971Art01}. A 290 MeV $^{40}$Ar beam from the Dubna 310 cm heavy-ion cyclotron bombarded a metallic $^{232}$Th. Reaction products were separated and identified with a magnetic spectrometer and a surface barrier silicon telescope. ``Apart from the nucleides already known, 17 new nucleides, namely: $^{29,30}$Mg, $^{31,32,33}$Al, $^{33,34,35,36}$Si, $^{35,36,37, 38}$P, $^{39,40}$S and $^{41,42}$Cl have been reliably detected.''

\subsubsection*{$^{41,42}$S}

In 1979 $^{41}$S and$^{42}$S were discovered by  Auger et al. in ``Observation of new nuclides $^{37}$Si, $^{40}$P, $^{41}$S, $^{42}$S produced in deeply inelastic reactions induced by $^{40}$Ar on $^{238}$U'' \cite{1979Aug01}.  A 263 MeV $^{40}$Ar beam from the Orsay ALICE facility bombarded a UF$_4$ target and reaction products were measured with a triple silicon solid state counter telescope. ``Four new neutron-rich nuclides, $^{37}$Si, $^{40}$P, $^{41-42}$S have been observed as a result of deep inelastic collisions. The nuclide identification combined two independent time of flight measurements as well as two ($\Delta$E $\times$ E) informations and was quite unambiguous.''

\subsubsection*{$^{43,44}$S}

The first observation of $^{43}$S and $^{44}$S was reported by Westfall et al. in ``Production of neutron-rich nuclides by fragmentation of 212-MeV/amu $^{48}$Ca'' in 1979 \cite{1979Wes01}. $^{48}$Ca ions (212 MeV/nucleon) from the Berkeley Bevalac were fragmented on a beryllium target. The fragments were selected by a zero degree spectrometer and identified in a telescope consisting of 12 Si(Li) detectors, 2 position-sensitive Si(Li) detectors, and a veto scintillator. ``In this letter, we present the first experimental evidence for the particle stability of fourteen nuclides $^{22}$N, $^{26}$F, $^{33,34}$Mg, $^{36,37}$Al, $^{38,39}$Si, $^{41,42}$P, $^{43,44}$S, and $^{44,45}$Cl produced in the fragmentation of 212-MeV/amu $^{48}$Ca.''

\subsubsection*{$^{45-47}$S}

Guillemaud-Mueller et al. announced the discovery of $^{45}$S, $^{46}$S, and $^{47}$S in the 1989 article ``Observation of new neutron rich nuclei $^{29}$F, $^{35,36}$Mg, $^{38,39}$Al, $^{40,41}$Si, $^{43,44}$P, $^{45-47}$S, $^{46-49}$Cl, and $^{49-51}$Ar from the interaction of 55 MeV/u $^{48}$Ca+Ta'' \cite{1989Gui01}. A 55~MeV/u $^{48}$Ca beam was fragmented on a tantalum target at GANIL and the projectile-like fragments were separated by the zero degree doubly achromatic LISE spectrometer. ``[The figure] represents the two-dimensional plot (energy loss versus time-of-flight) obtained under these conditions after 40~h integration time with an average intensity of 150 enA. The new species $^{35,36}$Mg, $^{38,39}$Al, $^{40,41}$Si, $^{43,44}$P, $^{45,46,47}$S, $^{46,47,48,49}$Cl, and $^{49,50,51}$Ar are clearly visible.''

\subsubsection*{$^{48}$S}

Lewitowicz et al. discovered $^{48}$S in the 1990 paper ``First observation of the neutron-rich nuclei $^{42}$Si, $^{45,46}$P, $^{48}$S, and $^{51}$Cl from the interaction of 44 MeV/u $^{48}$Ca + $^{64}$Ni'' \cite{1990Lew01}. A 44~MeV/u $^{48}$Ca beam was fragmented on a $^{64}$Ni  target at GANIL and the projectile-like fragments were separated by the zero degree doubly achromatic LISE spectrometer. ``The isotopes of $^{42}$Si, $^{45,46}$P, $^{48}$S, and $^{51}$Cl are identified for the first time.''

\subsection{Chlorine}\vspace{0.0cm}

The observation of 21 chlorine isotopes has been reported so far, including 2 stable, 5 proton-rich, and 14 neutron-rich isotopes. The proton dripline has been reached because as stated in \cite{2004Tho01} the particle identification plot in the 1986 paper by Langevin et al. clearly showed that $^{29}$Cl and $^{30}$Cl are beyond the dripline and have lifetimes shorter than the time-of-flight of 170~ns. According to the HFB-14 model \cite{2007Gor01}, $^{54}$Cl should be the last odd-odd particle stable neutron-rich nucleus while the odd-even particle stable neutron-rich nuclei should continue through $^{63}$Cl. Thus, about 8 isotopes have yet to be discovered corresponding to 28\% of all possible chlorine isotopes.

\begin{figure}
	\centering
	\includegraphics[scale=0.7]{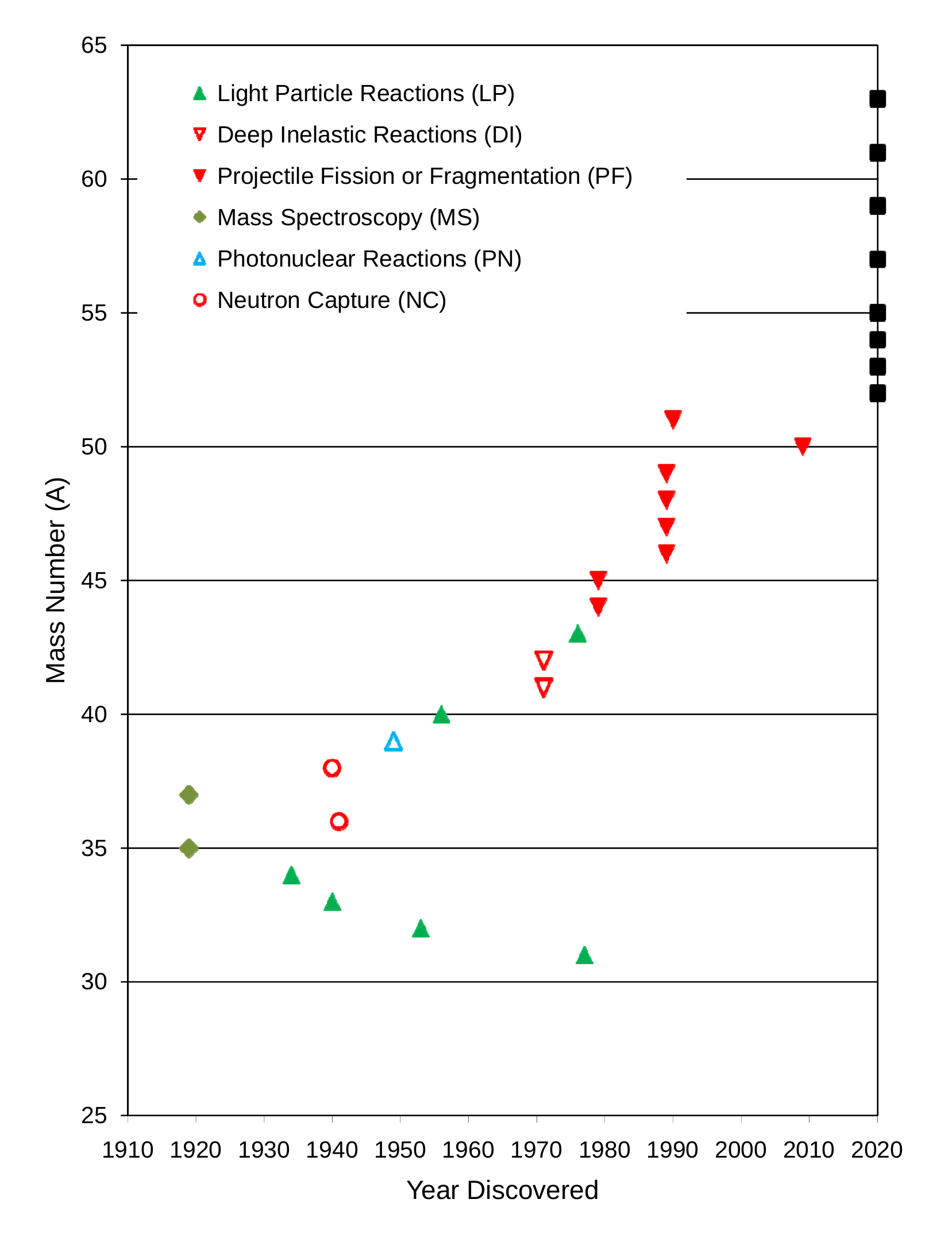}
	\caption{Chlorine isotopes as a function of time when they were discovered. The different production methods are indicated. The solid black squares on the right hand side of the plot are isotopes predicted to be bound by the HFB-14 model.}
\label{f:year-cl}
\end{figure}

Figure \ref{f:year-cl} summarizes the year of first observation for all chlorine isotopes identified by the method of discovery. The radioactive chlorine isotopes were produced using light-particle reactions (LP), deep-inelastic reactions (DI), neutron-capture reactions (NC), photo-nuclear reactions (PN), and projectile fragmentation of fission (PF). The stable isotopes were identified using mass spectroscopy (MS). Light particles are defined as incident particles with A$\le$4. The discovery of each chlorine isotope is discussed in detail and a summary is presented in Table 1.

\subsubsection*{$^{31}$Cl}

Benenson et al. observed $^{31}$Cl in 1977 as reported in ``Mass of $^{27}$P and $^{31}$Cl'' \cite{1977Ben01}. A $^{36}$Ar gas target was bombarded with a 70~MeV $^3$He beam and $^{31}$Cl was formed in a ($^3$He,$^8$Li) reaction. The ejectiles were measured with a double proportional counter and a thin plastic scintillator at the focal plane of a split-pole spectrograph. ``The final value for the Q value of $^{36}$Ar($^3$He,$^8$Li)$^{31}$Cl was 29180$\pm$50~keV which corresponds to a mass excess of $-$7070$\pm$50~keV.''

\subsubsection*{$^{32}$Cl}

In 1953 $^{32}$Cl was reported by Glass et al. in ``The short-lived radioisotopes P$^{28}$ and Cl$^{32}$'' \cite{1953Gla01}. Protons were accelerated to 20 MeV by the UCLA cyclotron and bombarded sulfur targets. $^{32}$Cl was produced in (p,n) charge exchange reactions and identified by measuring $\gamma$-rays with a NaI crystal. ``The half-life of the Cl$^{32}$ activity is 0.306$\pm$0.004 second, and in addition to positrons it emits gamma-radiation of energy 4.8$\pm$0.2~Mev.'' This half-life is in agreement with the currently adopted value of 298(1)~ms.

\subsubsection*{$^{33}$Cl}

Hoag reported the discovery of $^{33}$Cl in 1940 in ``The production and half-life of chlorine 33'' \cite{1940Hoa01}. High purity sulphur targets were bombarded with 8-MeV deuterons from the Berkeley 37'' cyclotron. The resulting activity was measured with an ionization chamber and Dershem electrometer and recorded on a kymograph. ``The record showed a decay curve which could be analyzed into two components of 2.5 min. and 2.8 sec. The former is P$^{30}$ formed in the known reaction of S$^{32}$+d$\rightarrow$P$^{30}$+$\alpha$. The short period gives an exponential decay over a factor of 100 in intensity. The saturation activities of the two periods are almost the same, which would rule out the possibility that the short one was due to a contamination or to any of the rare sulphur isotopes. The only other common type of reaction to be expected is S$^{32}$+d$\rightarrow$Cl$^{33}$+n. A (d,2n) reaction giving rise to Cl$^{32}$ can almost be ruled out on energetic grounds. We therefore conclude that the 2.8-sec. period is due to the decay of Cl$^{33}$ in the reaction Cl$^{33}\rightarrow$S$^{33}$+e$^+$.'' This half-life agrees with the currently adopted value of 2.511(3)~s.

\subsubsection*{$^{34}$Cl}

The first observation of $^{34}$Cl was reported by Frisch et al. in ``Induced radioactivity of sodium and phosphorus'' in 1934 \cite{1934Fri01}. A 1 mCi thorium B + C $\alpha$ source was used to irradiate phosphorus targets and the subsequent activity was measured with a Geiger-M\"uller counter. ``I have found that both sodium and phosphorus become active after $\alpha$-ray bombardment... Phosphorus (elementary red phosphorus) showed a very much longer lifetime. The half value period was found to be 40$\pm$5 minutes.... So for sodium and phosphorus the reactions would be $_{11}$Na$^{23}$ + $\alpha$ = $_{13}$Al$^{26}$ + neutron and $_{15}$P$^{31}$ + $\alpha$ = $_{17}$Cl$^{34}$ + neutron, respectively.'' This half-life is close to the presently accepted value of 32.00(4)~min for the isomeric state.

\subsubsection*{$^{35}$Cl}

The 1919 paper ``The constitution of the elements'' by Aston can be considered the discovery of $^{35}$Cl \cite{1919Ast01}. $^{35}$Cl was identified using the positive-ray mass spectrograph in Cambridge, England. ``The mass~spectra obtained when chlorine is present cannot be treated in detail here, but they appear to prove conclusively that this element consists of at least two isotopes of atomic weights 35 and 37.''

\subsubsection*{$^{36}$Cl}

Grahame and Walke reported the observation of $^{36}$Cl in the 1941 paper ``Preparation and properties of long-lived radio-chlorine'' \cite{1941Gra01}. ``Irradiation was carried out by allowing relatively large quantities (about a pound each) of sodium chlorate or of sodium perchlorate to stand in the neighborhood of the target holder of the Berkeley 37-inch cyclotron for periods of six months or more while the cyclotron was in use for other purposes.'' Activities were measured with a Lauritsen quartz fiber electroscope and a thin-walled counter. ``The emission of positrons, taken together with the fact that the familiar 37-minute radio-chlorine is known to be Cl$^{38}$, makes it reasonably certain that the new isotope is Cl$^{36}$ formed by the reaction Cl$^{35}$(n,$\gamma$)Cl$^{36}$.''

\subsubsection*{$^{37}$Cl}

The 1919 paper ``The constitution of the elements'' by Aston can be considered the discovery of $^{37}$Cl \cite{1919Ast01}. $^{37}$Cl was identified using the positive-ray mass spectrograph in Cambridge, England. ``The mass~spectra obtained when chlorine is present cannot be treated in detail here, but they appear to prove conclusively that this element consists of at least two isotopes of atomic weights 35 and 37.''

\subsubsection*{$^{38}$Cl}

In 1940 Kennedy and Seaborg reported the observation of $^{38}$Cl in``Isotopic identification of induced radioactivity by bombardment of separated isotopes; 37-minute Cl$^{38}$'' \cite{1940Ken01}. A previously reported 37-min half-life could be due to either $^{36}$Cl or $^{38}$Cl. An enriched HCl$^{35}$ and an ordinary HCl solution was activated with paraffin-slowed neutrons which were produced by beryllium bombardment with 16 MeV deuterons from the Berkeley 60-in. cyclotron. ``The lower intensity in the HCl$^{35}$ sample shows that this activity is to be assigned to Cl$^{38}$, formed as the result of neutron absorption by the heavier isotope Cl$^{37}$.'' The half-life agrees with the presently accepted value of 37.24(5)~min. The previous observations were produced in the reactions Cl(n,$\gamma$) \cite{1935Ama01}, Cl(d,p) \cite{1936Van01}, and K(n,$\alpha$) \cite{1937Hur01}.

\subsubsection*{$^{39}$Cl}

$^{39}$Cl was identified in 1949 by Haslam et al. in ``Confirmation of Cl$^{39}$ activity'' \cite{1949Has01}. The University of Saskatchewan betatron was used to irradiate argon at a betatron energy of 23 MeV. The resulting activities were measured with a thin-walled beta counter. ``The activity measured in the counting chamber, and thus due to the filtered argon, was then found to have a half-life of exactly 110 minutes, and the glass woolantimony filter carried a $\beta^-$-activity of 55.5$\pm$0.2 minutes. This is ascribed to the isotope Cl$^{39}$ produced in the reaction A$^{40}$($\gamma$,p)Cl$^{39}$. This isotope is listed in the table of Seaborg and Perlman \cite{1948Sea01} as having a half-life of one hour. This result is based on unpublished data.'' The half-life is in agreement with the presently accepted value of 55.6(2)~min.

\subsubsection*{$^{40}$Cl}

$^{40}$Cl was discovered by Morinaga as reported in ``Radioactive isotopes Cl$^{40}$ and Ga$^{74}$'' in 1956 \cite{1956Mor01}. Solid argon targets were irradiated with fast neutrons produced by bombarding a beryllium target with 10 MeV deuterons from the Purdue cyclotron. Gamma- and beta-rays were measured with a NaI scintillator and GM counter, respectively. ``From both gamma-ray measurements with a NaI scintillator and beta-ray measurements with a GM counter, the half-life of this new activity was found to be about 1.4 min. Since Cl$^{40}$ is the only unknown isotope which could be produced by irradiating argon and since moreover the energy of one of the gamma rays (1.46 Mev) coincides with the energy of the first excited state of A$^{40}$, this new activity is attributed to Cl$^{40}$.'' This half-life is consistent with the currently adopted value of 1.35(2)~min.

\subsubsection*{$^{41,42}$Cl}

Artukh et al. discovered $^{41}$Cl and $^{42}$Cl in the 1971 paper ``New isotopes $^{29,30}$Mg, $^{31,32,33}$Al, $^{33,34,35,36}$Si, $^{35,36,37,38}$P, $^{39,40}$S, and $^{41,42}$Cl produced in bombardment of a $^{232}$Th target with 290 MeV $^{40}$Ar ions'' \cite{1971Art01}. A 290 MeV $^{40}$Ar beam from the Dubna 310 cm heavy-ion cyclotron bombarded a metallic $^{232}$Th. Reaction products were separated and identified with a magnetic spectrometer and a surface barrier silicon telescope. ``Apart from the nucleides already known, 17 new nucleides, namely: $^{29,30}$Mg, $^{31,32,33}$Al, $^{33,34,35,36}$Si, $^{35,36,37, 38}$P, $^{39,40}$S and $^{41,42}$Cl have been reliably detected.''

\subsubsection*{$^{43}$Cl}

In 1976, Kashy et al. published the discovery of $^{43}$Cl in their paper ``Observation of highly neutron-rich $^{43}$Cl and $^{59}$Mn" \cite{1976Kas01}. A $^{48}$Ca target was bombarded with a 74 MeV $^{3}$He beam at the Michigan State University Cyclotron. $^{43}$Cl was produced in the reaction $^{48}$Ca($^3$He,$^8$B) and identified with an Enge split pole spectrograph. ``We report the observation and mass measurement of $^{43}$Cl and $^{59}$Mn by the ($^{3}$He, $^{8}$B), five-nucleon pickup reaction.''

\subsubsection*{$^{44,45}$Cl}

The first observation of $^{44}$Cl and $^{45}$Cl was reported by Westfall et al. in ``Production of neutron-rich nuclides by fragmentation of 212-MeV/amu $^{48}$Ca'' in 1979 \cite{1979Wes01}. $^{48}$Ca ions (212 MeV/nucleon) from the Berkeley Bevalac were fragmented on a beryllium target. The fragments were selected by a zero degree spectrometer and identified in a telescope consisting of 12 Si(Li) detectors, 2 position-sensitive Si(Li) detectors, and a veto scintillator. ``In this letter, we present the first experimental evidence for the particle stability of fourteen nuclides $^{22}$N, $^{26}$F, $^{33,34}$Mg, $^{36,37}$Al, $^{38,39}$Si, $^{41,42}$P, $^{43,44}$S, and $^{44,45}$Cl produced in the fragmentation of 212-MeV/amu $^{48}$Ca.''

\subsubsection*{$^{46-49}$Cl}

Guillemaud-Mueller et al. announced the discovery of $^{46}$Cl, $^{47}$Cl, $^{48}$Cl, and $^{49}$Cl in the 1989 article ``Observation of new neutron rich nuclei $^{29}$F, $^{35,36}$Mg, $^{38,39}$Al, $^{40,41}$Si, $^{43,44}$P, $^{45-47}$S, $^{46-49}$Cl, and $^{49-51}$Ar from the interaction of 55 MeV/u $^{48}$Ca+Ta'' \cite{1989Gui01}. A 55~MeV/u $^{48}$Ca beam was fragmented on a tantalum target at GANIL and the projectile-like fragments were separated by the zero degree doubly achromatic LISE spectrometer. ``[The figure] represents the two-dimensional plot (energy loss versus time-of-flight) obtained under these conditions after 40~h integration time with an average intensity of 150 enA. The new species $^{35,36}$Mg, $^{38,39}$Al, $^{40,41}$Si, $^{43,44}$P, $^{45,46,47}$S, $^{46,47,48,49}$Cl, and $^{49,50,51}$Ar are clearly visible.''

\subsubsection*{$^{50}$Cl}

$^{50}$Cl was discovered by Tarasov et al. in 2009 and published in ``Evidence for a change in the nuclear mass surface with the discovery of the most neutron-rich nuclei with 17 $\le$ Z $\le$ 25'' \cite{2009Tar01}. $^9$Be targets were bombarded with 132 MeV/u $^{76}$Ge ions accelerated by the Coupled Cyclotron Facility at the National Superconducting Cyclotron Laboratory at Michigan State University. $^{50}$Cl was produced in projectile fragmentation reactions and identified with a two-stage separator consisting of the A1900 fragment separator and the S800 analysis beam line. ``The observed fragments include fifteen new isotopes that are the most neutron-rich nuclides of the elements chlorine to manganese ($^{50}$Cl, $^{53}$Ar, $^{55,56}$K, $^{57,58}$Ca, $^{59,60,61}$Sc, $^{62,63}$Ti, $^{65,66}$V, $^{68}$Cr, $^{70}$Mn).''

\subsubsection*{$^{51}$Cl}

Lewitowicz et al. discovered $^{51}$Cl in the 1990 paper ``First observation of the neutron-rich nuclei $^{42}$Si, $^{45,46}$P, $^{48}$S, and $^{51}$Cl from the interaction of 44 MeV/u $^{48}$Ca + $^{64}$Ni'' \cite{1990Lew01}. A 44~MeV/u $^{48}$Ca beam was fragmented on a $^{64}$Ni  target at GANIL and the projectile-like fragments were separated by the zero degree doubly achromatic LISE spectrometer. ``The isotopes of $^{42}$Si, $^{45,46}$P, $^{48}$S, and $^{51}$Cl are identified for the first time.'' The observation was later questioned by Tarasov et al.: ``While not conclusive, the previous identification of this isotope may have been masked by the presence of the hydrogenlike ion $^{48}$Cl$^{16+}$ produced at the same time.'' \cite{2009Tar01}.

\subsection{Argon}\vspace{0.0cm}

The chemical symbol for argon was A until it was changed to Ar in 1957. The observation of 23 argon isotopes has been reported so far, including 3 stable, 7 proton-rich, and 13 neutron-rich isotopes. No specific searches for the existence of $^{30}$Ar have been reported and so it could potentially still be observed \cite{2004Tho01}. The latest mass evaluation predicts the one-proton separation energy of $^{30}$Ar to be 350(360)~keV. According to the HFB-14 model \cite{2007Gor01}, $^{55}$Ar should be the last odd-even particle stable neutron-rich nucleus while the even-even particle stable neutron-rich nuclei should continue through $^{66}$Ar. Thus, about 8 isotopes have yet to be discovered corresponding to 16\% of all possible argon isotopes.

\begin{figure}
	\centering
	\includegraphics[scale=0.7]{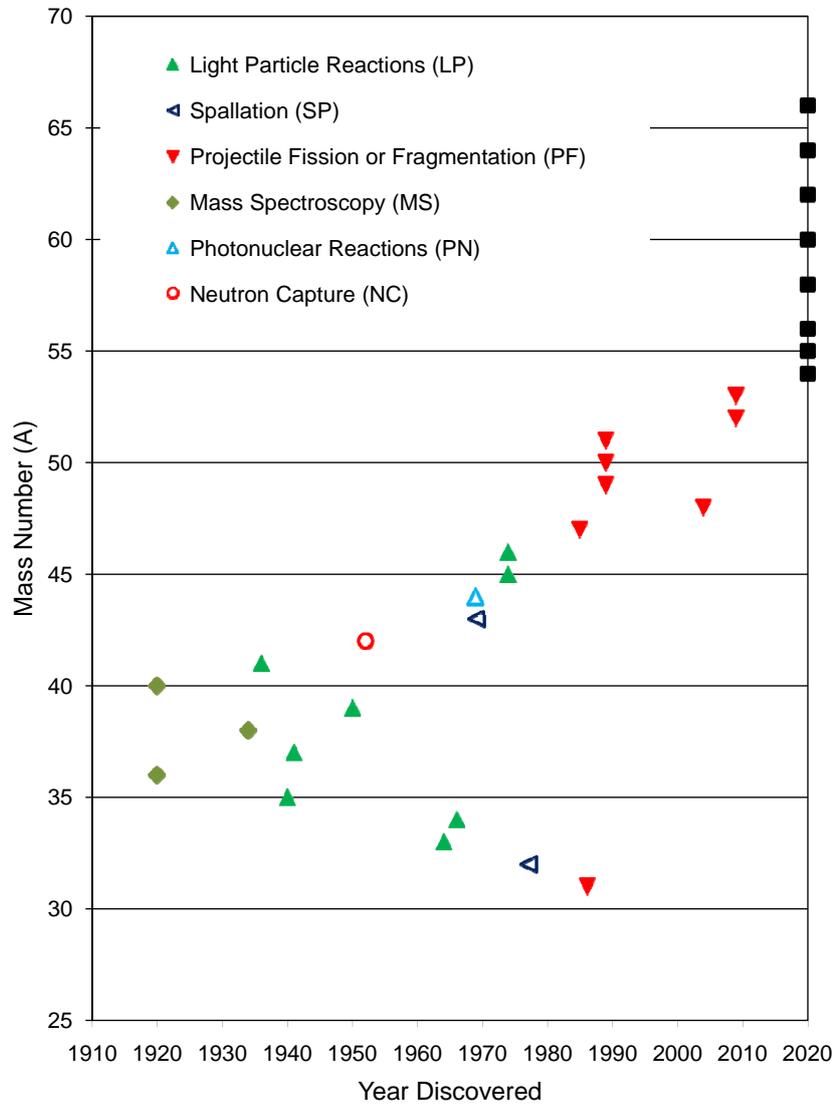}
	\caption{Argon isotopes as a function of time when they were discovered. The different production methods are indicated. The solid black squares on the right hand side of the plot are isotopes predicted to be bound by the HFB-14 model.}
\label{f:year-ar}
\end{figure}

Figure \ref{f:year-ar} summarizes the year of first observation for all argon isotopes identified by the method of discovery. The radioactive argon isotopes were produced using light-particle reactions (LP), spallation (SP), neutron-capture reactions (NC), photo-nuclear reactions (PN), and projectile fragmentation of fission (PF). The stable isotopes were identified using mass spectroscopy (MS). Light particles are defined as incident particles with A$\le$4. The discovery of each argon isotope is discussed in detail and a summary is presented in Table 1.

\subsubsection*{$^{31}$Ar}

$^{31}$Ar was first reported in 1986 by Langevin et al. in ``Mapping of the proton drip-line up to Z = 20: Observation of the T$_z$=$-$5/2 series $^{23}$Si, $^{27}$S, $^{31}$Ar, and $^{35}$Ca'' \cite{1986Lan01}. A 77.4~MeV/u $^{40}$Ca beam was fragmented on a nickel target at GANIL and the projectile-like fragments were separated by the zero degree doubly achromatic LISE spectrometer. The isotopes were identified by measuring energy loss and time-of-flight. ``The bidimensional plot (see [the figure]) of $\sqrt{\Delta}$/t.o.f. (i.e. Z) versus t.o.f (i.e. A/Z) was inspected on-line to calibrate the particle identification... [The figure] shows the same bidimensional representation after 14 hours of integration time. The T$_z$ series $^{23}$Si, $^{27}$S, $^{31}$Ar, and $^{35}$Ca clearly becomes visible.''

\subsubsection*{$^{32}$Ar}

Hagberg et al. discovered $^{32}$Ar in the 1977 paper ``Decay of a T$_z$=$-$2 nucleus: Argon-32'' \cite{1977Hag01}. 600 MeV protons from the CERN synchrocyclotron bombarded a vanadium target. $^{32}$Ar was produced in spallation reactions and identified with the ISOLDE electromagnetic isotope separator. Beta-delayed protons were measured with a silicon surface barrier counter. ``The $\beta$-delayed proton spectrum observed for $^{32}$Ar is shown in [the figure]. Only one peak is evident,
with a laboratory energy of 3350.5$\pm$5.0 keV. The time decay of the peak yields a half-life for $^{32}$Ar of 75$^{+70}_{-30}$~msec.'' This half-life agrees with the presently adopted value of 98(2)~ms.

\subsubsection*{$^{33}$Ar}

The discovery of $^{33}$Ar was reported in 1964 by Reeder et al. in ``New delayed-proton emitters: Ti$^{41}$, Ca$^{37}$, and Ar$^{33}$'' \cite{1964Ree01}. The Brookhaven 60-in. cyclotron bombarded gaseous H$_2$S and solid sulfur targets with $^3$He at a maximum energy of 31.8 MeV. Proton spectra were measured by two surface barrier detectors. ``The excitation function observed for Ca$^{37}$ has a threshold at 20$\pm$2 MeV which is consistent with the predicted threshold of 19.4 MeV for the (He$^3$,2\textit{n}) reaction. ``Three new nuclides, Ti$^{41}$, Ca$^{37}$, and Ar$^{33}$, have been observed to be delayed proton emitters of the type that undergo beta decay to proton unstable states of daughter nuclei.'' The reported half-life of 182(5)~ms is consistent with the currently adopted value of 173.0(20)~ms. Independently, Hardy and Verrall reported a 178(10)~ms only two weeks later \cite{1965Har01}.

\subsubsection*{$^{34}$Ar}

Miller and Kavanagh reported the observation of $^{34}$Ar in the 1966 paper ``Decay of $^{34}$Ar'' \cite{1966Mil01}. A 10~MeV $^3$He from the ONR-CIT tandem accelerator bombarded a Sb$_2$S$_3$ target and $^{34}$Ar was produced in the reaction $^{32}$S($^3$He,n). The resulting activities were measured with a NaI(Tl) crystal. ``The bottom spectrum of [the figure] shows the difference in yields between the first and second seconds after beam turn-off. Only three statistically significant peaks remain. They are located at energies (in MeV) of 0.51, 0.67 and 1.02. The 0.51 and 1.02 peaks are due to annihilation quanta,
singly and in random coincidence. The remaining peak, at 0.67 MeV, was found to have a half life of 1.2$\pm$0.3~s and is attributed to the decay of $^{34}$Ar to the 0.67 MeV state of $^{34}$Cl.'' This half-life agrees with the presently adopted value of 845(3)~ms.

\subsubsection*{$^{35}$Ar}

King and Elliott identified $^{35}$Ar in ``Short-lived radioactivities of $_{14}$Si$^{27}$, $_{16}$S$^{31}$, and $_{18}$A$^{35}$'' in 1940 \cite{1940Kin01}. Sulfur targets were bombarded with 16~MeV $\alpha$ particles and the resulting activities were measured with a multiple Geiger counter circuit. ``In an attempt to extend the well-known series of radioactive elements characterized by the formula Z $-$ N = 1, the following new reactions have been observed:... Reaction: $_{16}$S$^{32}$($\alpha$,n)$_{18}$A$^{35}$; Half-life: 1.91~s.'' This half-life agrees with the presently accepted value of 1.775(4)~s.

\subsubsection*{$^{36}$Ar}

In 1920 $^{36}$Ar was first measured by Aston in ``The constitution of the elements'' \cite{1920Ast01}. The isotope was identified in a mass spectrometer at Cambridge, England. ``Argon (atomic weight 39.88 Ramsay; 39.91 Leduc) gives a very strong line exactly at 40, with double charge at 20 and triple charge at 13$\frac{1}{3}$. The last line, being closely bracketed by known reference lines at 13 and, 14, provides very trustworthy values. At first this was thought to be its only constituent, but further photographs showed an associated faint line at 36. This has not yet been proved an element by double and triple charges, as the probable presence of OH$_2$ and the certain presence of C prevent this, but other lines of reasoning make it extremely probable that this is a true isotope, the presence of which to the extent of 3 per cent. is enough to account for the fractional atomic weight quoted.''

\subsubsection*{$^{37}$Ar}

Weimer et al. reported the first observation of $^{37}$Ar in the 1941 article ``Radioactive argon A$^{37}$'' \cite{1941Wei01}. A variety of reactions were used to produce $^{37}$Ar and the resulting activities were measured with an ionization chamber connected to a Wulf bifilar electrometer. ``An artificially radioactive gas has been produced by bombarding solid samples containing potassium, chlorine, calcium, or sulfur with appropriate nuclear particles. The radioactivity has been observed for three months and is found to have a single decay period of 34 days... In view of the method by which the present activity has
been produced, we assign it to A$^{37}$.'' This half-life agrees with the currently accepted value of 35.04(4)~d.

\subsubsection*{$^{38}$Ar}

$^{38}$Ar was discovered in 1934 by Zeeman and de Gier as reported in ``A new isotope of argon'' \cite{1934Zee01}. Argon gas was examined in a mass spectrograph. ``Between the parabolas for the isotopes A$^{40}$ and A$^{36}$ always a parabola for the mass 38 was obtained. The intensity of the 38 parabola relatively to that of the two other ones remained unchanged by diluting with O$_2$N$_2$, and other gases... Atoms of mass 38 were till now unknown. We therefore infer, that 38 is really due to a new isotope of argon.''

\subsubsection*{$^{39}$Ar}

In the 1950 article ``Argon$^{39}$ beta-spectrum'' Brosi et al. described the observation of $^{39}$Ar \cite{1950Bro01}. Potassium salt was irradiated with neutrons in nuclear reactors. Activities were measured with a proportional counter and a NaI(Tl) detector. ``This new long-lived argon isotope is presumably A$^{39}$ formed by an (n,p) reaction on K$^{39}$. An attempt to find the 4-min. activity previously assigned to A$^{39}$ was unsuccessful.'' The 1940 table of isotopes assigns a 4 min half-life to $^{39}$Ar \cite{1940Liv01} quoting a 1937 paper by Pool et al. However, although Pool reported a 4 min half-life observed following the irradiation of potassium with neutrons, they did not assign the activity to a specific isotope \cite{1937Poo01}.

\subsubsection*{$^{40}$Ar}

In 1920 $^{40}$Ar was first measured by Aston in ``The constitution of the elements'' \cite{1920Ast01}. The isotope was identified in a mass spectrometer at Cambridge, England. ``Argon (atomic weight 39.88 Ramsay; 39.91 Leduc) gives a very strong line exactly at 40, with double charge at 20 and triple charge at 13$\frac{1}{3}$. The last line, being closely bracketed by known reference lines at 13 and, 14, provides very trustworthy values.''

\subsubsection*{$^{41}$Ar}

The discovery of $^{41}$Ar was reported in 1936 by Snell in ``Radioactive argon'' \cite{1936Sne01}. Argon gas was bombarded with 3 MeV deuterons from the Lawrence and Livingston magnetic resonance accelerator. Beta-ray absorption and decay spectra as well as $\gamma$-ray were recorded. ``When bombarded with high speed deuterons, argon gas is found to yield a radioactive product which emits negative electrons, and decays with a period of 110$\pm$1 minutes. Chemical tests show that the activity is due to an isotope of argon, and the reaction involved is doubtless A$^{40}$+H$^2$=A$^{41}$+H$^1$.'' This half-life agrees with the presently adopted value of 109.61(4)~min.

\subsubsection*{$^{42}$Ar}

In 1952 Katcoff reported the observation of $^{42}$Ar in ``Thermal neutron capture cross section of A$^{40}$ and observation of A$^{42}$'' \cite{1952Kat01}. Pure argon gas was irradiated with neutrons from the Brookhaven pile and $^{42}$Ar was formed by two successive neutron captures. The resulting activities were measured with a proportional counter. ``No attempt was made to detect the A$^{42}$ radiations directly because of the greatly preponderant activity of A$^{39}$. Rather, the A$^{42}$ was detected by successive extractions of its 12.5-hr K$^{42}$ daughter...  In 13 extractions over a period of 400 days, the corrected activity of A$^{42}$ showed no apparent decrease; consideration of the possible errors indicates that it could not have gone down by more than 20 percent. This sets a lower limit of 3.5 years on the half~life of A$^{42}$...'' This limit is consistent with the present value of 32.9(11)~y.

\subsubsection*{$^{43}$Ar}

Hansen et al. reported the first observation of $^{43}$Ar in the paper ``Decay characteristics of short-lived radio-nuclides studied by on-line isotope separator techniques'' in 1969 \cite{1969Han01}. 600 MeV protons from the CERN synchrocyclotron bombarded a TiO$_2$(H$_2$O)$_x$ target and argon isotopes were separated using the ISOLDE facility. Electron capture, $\beta$- and $\gamma$-rays were measured. The paper summarized the ISOLDE program and did not contain details about the individual nuclei other than in tabular form. The measured half-life of 5.35(15)~min agrees with the presently adopted value of 5.37(6)~min. Less than 6 months later Larson and Gordon independently reported a half-life of 6.5(18)~min \cite{1969Lar01}.

\subsubsection*{$^{44}$Ar}

Larson and Gordon reported the observation of $^{44}$Ar in the 1969 paper ``Production and decay of $^{43}$Ar and $^{44}$Ar''  \cite{1969Lar01}. Enriched $^{48}$Ca targets were irradiated with bremsstrahlung from the NRL Linac. The resulting activity was measured with a Ge(Li) detector. ``These data are compared with growth curves computed for $^{44}$Ar decaying with half-life of 14 min into $^{44}$K, which has a 22 min half-life. A curve computed for a six min activity decaying into $^{44}$K (T$_{1/2}$ = 22 min) is also shown. It can be seen that the measured values from [the figure] agree with a 14 min half-life for $^{44}$Ar.'' This half-life is consistent with the currently adopted value of 11.87(5)~min.

\subsubsection*{$^{45,46}$Ar}

Jelley et al. discovered $^{45}$Ar and $^{46}$Ar in 1974 as described in ``Masses for $^{43}$Ar and the new isotopes $^{45}$Ar and $^{46}$Ar'' \cite{1974Jel01}. Enriched $^{48}$Ca targets were bombarded with 77.7 MeV $\alpha$-particles and 80.1 MeV $^6$Li from the Berkeley 88-in. cyclotron to form $^{45}$Ar and $^{46}$Ar, respectively. The ejectiles were measured with a counter telescope. ``By also detecting $^7$Be nuclei from the $^{48}$Ca($\alpha$,$^7$Be)$^{45}$Ar reaction (Q$^\sim -$28 MeV), excited states in $^{45}$Ar and the mass of this new isotope were determined. Similarly, since the feasibility of employing the ($^6$Li,$^8$B) two-proton transfer reaction as a means of studying neutron-rich nuclei has been demonstrated, the $^{48}$Ca($^8$Li,$^8$B)$^{46}$Ar reaction (Q$^\sim -$23 MeV) was used to establish the mass of $^{46}$Ar.''

\subsubsection*{$^{47}$Ar}

Guillemaud-Mueller et al. announced the discovery of $^{47}$Ar in the 1985 article ``Production and identification of new neutron-rich fragments from 33~MeV/u $^{86}$Kr beam in the 18$\leq$Z$\leq$27 region} \cite{1985Gui01}.  At GANIL in Caen, France, a 33~MeV/u $^{86}$Kr beam was fragmented and the fragments were separated by the triple-focusing analyser LISE.  ``Each particle is identified by an event-by-event analysis.  The mass A is determined from the total energy and the time of flight, and Z by the $\Delta$E and E measurements... In addition to that are identified the following new isotopes:  $^{47}$Ar, $^{57}$Ti, $^{59,60}$V, $^{61,62}$Cr, $^{64,65}$Mn, $^{66,67,68}$Fe, $^{68,69,70}$Co.'' Only 3 days later Benenson et al. reported a mass measurement of $^{47}$Ar \cite{1985Ben01}.

\subsubsection*{$^{48}$Ar}

In 2004 Gr\'evy et al. identified $^{48}$Ar in ``Beta-decay half-lives at the N = 28 shell closure'' \cite{2004Gre01}. A 60~MeV/u $^{48}$Ca beam was fragmented on a beryllium target at GANIL and the projectile-like fragments were separated by the zero degree doubly achromatic LISE3 spectrometer. Beta-particles were measured with two plastic scintillators correlated with the implantation of the fragments in a double-sided Si-strip detector. ``We report here on the measurements of the $\beta$-decay half-lives of nuclei between $^{36}$Mg (N = 24) and $^{48}$Ar (N = 30).'' The measured 475(40)~ms half-life corresponds to the currently adopted value. Guillemaud-Mueller et al. reported the observation of even more neutron-rich argon isotopes ($^{49-51}$Ar) earlier, but did not mention or show any evidence for $^{48}$Ar referring to an internal report \cite{1988Zha01}.

\subsubsection*{$^{49-51}$Ar}

Guillemaud-Mueller et al. announced the discovery of $^{49}$Ar, $^{50}$Ar, and $^{51}$Ar in the 1989 article``Observation of new neutron rich nuclei $^{29}$F, $^{35,36}$Mg, $^{38,39}$Al, $^{40,41}$Si, $^{43,44}$P, $^{45-47}$S, $^{46-49}$Cl, and $^{49-51}$Ar from the interaction of 55 MeV/u $^{48}$Ca+Ta'' \cite{1989Gui01}. A 55~MeV/u $^{48}$Ca beam was fragmented on a tantalum target at GANIL and the projectile-like fragments were separated by the zero degree doubly achromatic LISE spectrometer. ``[The figure] represents the two-dimensional plot (energy loss versus time-of-flight) obtained under these conditions after 40~h integration time with an average intensity of 150 enA. The new species $^{35,36}$Mg, $^{38,39}$Al, $^{40,41}$Si, $^{43,44}$P, $^{45,46,47}$S, $^{46,47,48,49}$Cl, and $^{49,50,51}$Ar are clearly visible.''

\subsubsection*{$^{52,53}$Ar}

$^{52}$Ar and $^{53}$Ar were discovered by Tarasov et al. in 2009 and published in ``Evidence for a change in the nuclear mass surface with the discovery of the most neutron-rich nuclei with 17 $\le$ Z $\le$ 25'' \cite{2009Tar01}. $^9$Be targets were bombarded with 132 MeV/u $^{76}$Ge ions accelerated by the Coupled Cyclotron Facility at the National Superconducting Cyclotron Laboratory at Michigan State University. $^{52}$Ar and $^{53}$Ar were produced in projectile fragmentation reactions and identified with a two-stage separator consisting of the A1900 fragment separator and the S800 analysis beam line. ``The observed fragments include fifteen new isotopes that are the most neutron-rich nuclides of the elements chlorine to manganese ($^{50}$Cl, $^{53}$Ar, $^{55,56}$K, $^{57,58}$Ca, $^{59,60,61}$Sc, $^{62,63}$Ti, $^{65,66}$V, $^{68}$Cr, $^{70}$Mn).'' $^{52}$Ar was not specifically mentioned as a new observation because of the previous publication in a conference abstract \cite{2008Man02}, however, it is clearly visible in the particle identification plot of the measured atomic number Z versus the calculated function N$-$Z.

\subsection{Potassium}\vspace{0.0cm}

The observation of 22 potassium isotopes has been reported so far, including 3 stable, 4 proton-rich, and 15 neutron-rich isotopes. The proton dripline has been reached because as stated in \cite{2004Tho01} the particle identification plot in the 1986 paper by Langevin et al. clearly showed that $^{33}$K and $^{34}$K are beyond the dripline and have lifetimes shorter than the time-of-flight of 170~ns. According to the HFB-14 model \cite{2007Gor01}, $^{58}$K should be the last odd-odd particle stable neutron-rich nucleus while the odd-even particle stable neutron-rich nuclei should continue through $^{67}$K. Thus, about 7 isotopes have yet to be discovered corresponding to 24\% of all possible potassium isotopes.

\begin{figure}
	\centering
	\includegraphics[scale=0.7]{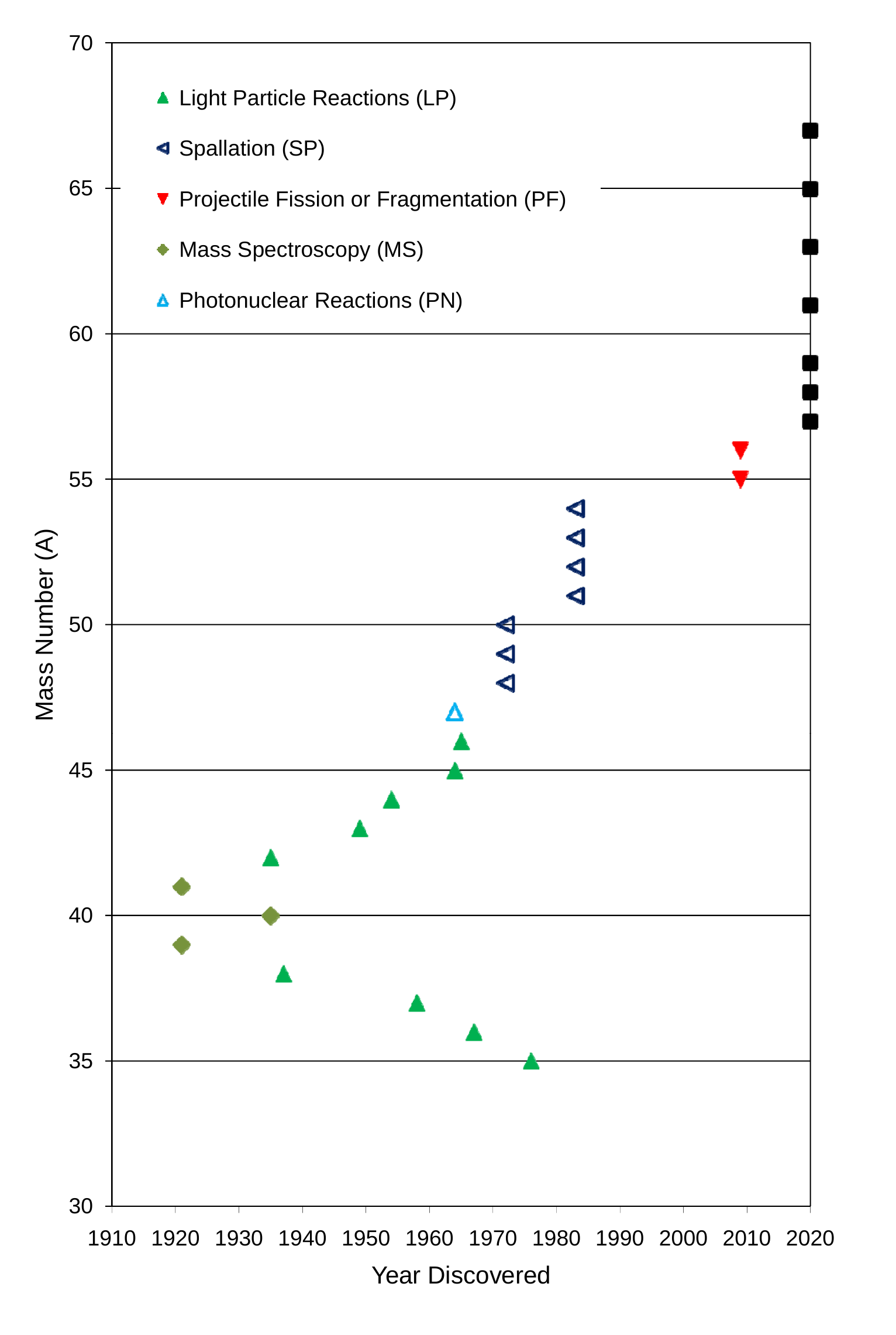}
	\caption{Potassium isotopes as a function of time when they were discovered. The different production methods are indicated. The solid black squares on the right hand side of the plot are isotopes predicted to be bound by the HFB-14 model.}
\label{f:year-k}
\end{figure}

Figure \ref{f:year-k} summarizes the year of first observation for all potassium isotopes identified by the method of discovery. The radioactive potassium isotopes were produced using light-particle reactions (LP), spallation (SP), photo-nuclear reactions (PN), and projectile fragmentation of fission (PF). The stable isotopes were identified using mass spectroscopy (MS). Light particles are defined as incident particles with A$\le$4. The discovery of each potassium isotope is discussed in detail and a summary is presented in Table 1.

\subsubsection*{$^{35}$K}

Benenson et al. discovered $^{35}$K as reported in the 1976 paper ``Mass of $^{35}$K'' \cite{1976Ben01}. An enriched $^{40}$Ca target was bombarded with 73.7 and 75.8~MeV $^3$He beams from the Michigan State cyclotron. $^{35}$K was produced in the reaction $^{40}$Ca($^3$He,$^8$Li) and identified by detecting the ejectiles with an Enge split pole spectrograph. ``The Q value for the reaction was found to be $-$29.693$\pm$0.020 MeV and the mass excess to be $-$11.170$\pm$0.020 MeV. Excited states of $^{35}$K were found at 1.56 and 2.69 MeV.''

\subsubsection*{$^{36}$K}

The existence of $^{36}$K was shown by Berg et al. in 1967 in ``$^{36}$K decay and T=1 analog in $^{36}$Ar'' \cite{1967Ber01}. An enriched $^{36}$Ar gas target was bombarded with 23 MeV protons from the Michigan State sector-focused cyclotron and $^{36}$K was formed in the (p,n) charge exchange reaction. Decay curves and $\gamma$-ray spectra were measured with a Ge(Li) detector. ``Values of the $^{36}$K half-life were calculated from such data by comparing the yield in successive spectra of each $\gamma$ ray identified as a transition in $^{36}$Ar. The resulting value for the $^{36}$K half-life after correction for the decrease in analyzer dead time in successive spectra was 0.265 sec, with a standard deviation of 0.025 sec.'' This half-life agrees with the presently adopted value of 342(2)~ms.

\subsubsection*{$^{37}$K}

In 1958 Sun and Wright identified $^{37}$K in ``Radionuclide K$^{37}$ \cite{1958Sun01}. A 12.8 MeV proton beam from the UCLA 20-MV synchrocyclotron bombarded a natural calcium target. Resulting activities were measured with a stilbene crystal scintillation counter and the half-life was determined with a Sanborn Twin-Viso recorder. ``The half-life as measured on the Sanborn recorder is T$_{1/2}$= 1.2$\pm$0.12 sec.'' This half-life agrees with the currently accepted value of 1.226(7)~s. A 1.2~s half-life had previously been assigned to $^{37}$K produced in the reaction $^{39}$K($\gamma$,2n) \cite{1951Bol01}, however, later this observation was assigned to be the 0.95~s isomeric state in $^{38}$K populated in the ($\gamma$,n) reaction \cite{1953Sta01}.

\subsubsection*{$^{38}$K}

Hurst and Walke reported the observation of $^{38}$K in ``The induced radioactivity of potassium'' in 1937 \cite{1937Hur01}. Lithium chloride was bombarded with 11 MeV $\alpha$ particles and $^{38}$K was formed in the reaction $^{35}$Cl($\alpha$,n). Decay and absorption curves as well as $\gamma$-ray spectra were measured following chemical separation. ``The precipitate had a strong activity decaying to half-value in 7.7$_5\pm0.1_5$ minutes as shown in [the figure]. The particles emitted were positrons having a maximum energy as determined by Feather's rule from the thickness of aluminum required to stop them of 2 Mev.'' This half-life is in agreement with the presently adopted value of 7.636(18)~min.

\subsubsection*{$^{39}$K}

The discovery of stable $^{39}$K was reported by Aston in his 1921 paper ``The constitution of the alkali metals'' \cite{1931Ast03}. The positive anode ray method was used to identify $^{39}$K with the Cavendish mass spectrograph. ``Potassium (atomic weight 39.l0) gives a strong line at 39 and a very weak companion at 41. These are integers within about a quarter of a unit compared with sodium 23. The relative intensities of the lines are not inconsistent with the accepted atomic weight. Potassium therefore probably consists of two isotopes 39 and 4l.''

\subsubsection*{$^{40}$K}

Nier discovered $^{40}$K in the 1935 paper ``Evidence for the existence of an isotope of potassium of mass 40'' \cite{1935Nie01}. Potassium was introduced into a mass spectrograph. ``[The figure] shows the interesting portion of one of many mass spectrographic analyses of the region around m/e=40. As may be seen there is a very definite peak due to an ion with an m/e value of 40. The rising portions on either side of this peak are the feet of the very much larger peaks due to K$^{39}$ and K$^{41}$.''

\subsubsection*{$^{41}$K}

The discovery of stable $^{41}$K was reported by Aston in his 1921 paper ``The constitution of the alkali metals'' \cite{1931Ast03}. The positive anode ray method was used to identify $^{41}$K with the Cavendish mass spectrograph. ``Potassium (atomic weight 39.l0) gives a strong line at 39 and a very weak companion at 41. These are integers within about a quarter of a unit compared with sodium 23. The relative intensities of the lines are not inconsistent with the accepted atomic weight. Potassium therefore probably consists of two isotopes 39 and 4l.''

\subsubsection*{$^{42}$K}

In 1935 Hevesy identified $^{42}$K in ``Natural and artificial radioactivity of potassium'' \cite{1935Hev03}. Scandium oxide was irradiated with neutrons from a beryllium-radium source and $^{42}$K was formed in the reaction $^{45}$Sc(n,$\alpha$). Beta-rays were measured following chemical separation. ``The scandium oxide was dissolved in hydrochloric acid and, after the addition of 0.15 gm. of sodium chloride and the same amount of calcium chloride, precipitated with ammonia. The calcium present in the filtrate was removed as oxalate and found to be inactive. The remaining sodium chloride, however, was found to be active and to contain the potassium isotopes looked for. This decayed with a period of about 16 hours, emitting very hard $\beta$-rays of approximately 1.2 million e.v.'' This half-life is consistent with the currently accepted value of 12.360(12)~h. Previously, Amaldi et al. had observed the 16~h half-life without a mass assignment \cite{1935Ama01}. Hurst and Walker reported a more accurate half-life of 12.4(2)~h questioning the half-life measurement by Hevesy \cite{1937Hur01}. However, Hurst and Walker acknowledge the population of $^{42}$K in the $^{45}$Sc(n,$\alpha$) and thus we credit Hevesy for the discovery of $^{42}$K.

\subsubsection*{$^{43}$K}

$^{43}$K was observed in 1949 by Overstreet et al. in the article ``Evidence for a new isotope of potassium'' \cite{1949Ove01}. Argon was bombarded with 40 MeV $\alpha$ particles and the subsequent activity was measured with a Lauritsen electroscope following chemical separation. ``A study of the possible isotopes resulting from the bombardment of argon with alpha-particles favors the assignment of K$^{43}$ to the 22.4 hour potassium isotope. The reaction would be A$^{40}$($\alpha$,p)K$^{43}$. This choice is further substantiated by the fact that the new isotope is a beta-emitter.'' This half-life agrees with the presently adopted value of 22.3(1)~h.

\subsubsection*{$^{44}$K}

Cohen identified $^{44}$K in the 1954 paper ``Potassium-44'' \cite{1954Coh02}. Natural calcium and enriched $^{44}$Ca was irradiated with neutrons which were produced by bombarding beryllium with 22 MeV protons. Gamma- and beta-ray spectra were measured following chemical separation. ``Potassium-44 was produced by an (n,p) reaction on calcium and found to decay by negatron and gamma emission with a 22.0$\pm$0.5 minute half-life. The identification was ascertained by comparison of yields from normal and isotopically enriched calcium, cross section measurements, chemical processing, and investigation of impurity effects.'' This half-life agrees with the currently accepted value of 22.13(19)~min. Previously, Walke had assigned a half-life of 18(1)~min to either $^{43}$K and $^{44}$K \cite{1937Wal01}.

\subsubsection*{$^{45}$K}

Morinaga and Wolzak discovered $^{45}$K in 1964 as reported in ``Potassium 45'' \cite{1964Mor01}. An enriched Ca$^{48}$CO$_3$ target was irradiated with 52 MeV $\alpha$-particles. Gamma-ray spectra were measured with a NaI(Tl) detector following chemical separation. ``Assignment of this activity to K$^{45}$ is most unambiguously made from its gamma spectrum. The energy of one of the most intense gamma ray (0.175 MeV) corresponds to the first excited state energy of Ca$^{45}$ and that of a 1.7 MeV gamma ray can be ascribed to the gamma ray from the 1.9 MeV state to the first excited state. The total decay energy of 4.0 MeV, which results from the decay scheme, also supports the assignment of the 20 min, activity to K$^{45}$.'' This half-life agrees with the currently adopted value of 17.3(6)~min.

\subsubsection*{$^{46}$K}

In the 1965 paper ``New isotope K$^{46}$ produced with the Ca$^{48}$(d,$\alpha$)K$^{46}$ reaction'' Marinov and Erskine reported the observation of $^{46}$K \cite{1965Mar01}. Enriched $^{48}$Ca targets were bombarded with 12 MeV deuterons from the Argonne Van de Graaff accelerator forming $^{46}$K in the (d,$\alpha$) reaction. The ejectiles were identified with a broad range magnetic spectrograph. ``The ground-state Q value in the Ca$^{48}$(d,$\alpha$)K$^{46}$ reaction was measured to be 1.915$\pm$0.015~MeV.''

\subsubsection*{$^{47}$K}

In 1964, Kuroyanagi et al. discovered $^{47}$K as described in ``Potassium-47'' \cite{1964Kur01}. Natural calcium and enriched $^{48}$Ca targets were irradiated with 23 MeV bremsstrahlung from the JAERI linac. Gamma- and beta-ray spectra were measured following chemical separation with a NaI(Tl) crystal and a plastic scintillator, respectively. ``From purely radiochemical considerations a new activity of 17.5$\pm$0.3 sec half-life is assigned to $^{47}$K.'' This half-life agrees with the presently adopted value of 17.50(24)~s.

\subsubsection*{$^{48-50}$K}

In 1972 Klapisch et al. reported the first observation of $^{48}$K, $^{49}$K, and $^{50}$K in ``Half-life of the new isotope $^{32}$Na; Observation of $^{33}$Na and other new isotopes produced in the reaction of high-energy protons on U'' \cite{1972Kla01}. Uranium targets were bombarded with 24 GeV protons from the CERN proton synchrotron. $^{48}$K, $^{49}$K, and $^{50}$K were identified by on-line mass spectrometry and decay curves were measured. ``Following the same procedure as for Na, the isotopes $^{48}$K, $^{49}$K, and $^{50}$K were found. However, their half-lives were not short compared with the diffusion time, and hence could not be determined.''

\subsubsection*{$^{51-54}$K}

Langevin et al. is credited with the discovery of $^{51}$K, $^{52}$K, $^{53}$K, and $^{54}$K in 1983 in ``$^{53}$K, $^{54}$K And $^{53}$Ca: Three new neutron rich isotopes'' \cite{1983Lan01}. Iridium was fragmented by 10 GeV protons from the CERN synchrotron to produce neutron rich potassium isotopes, which then decayed into calcium isotopes. Neutrons were measured in coincidence with $\beta$-rays after the potassium was mass separated. ``This work gives evidence for three new K and Ca isotopes and provides further information on half-lives and P$_n$ values.'' Half-lives of 365(5)~ms for $^{51}$K, 105(5)~ms for $^{52}$K, 30(5)~ms for $^{53}$K, and 10(5)~ms for $^{54}$K were reported and correspond to the presently adopted values. The observation of $^{51}$K and $^{52}$K was not considered a discovery of new isotopes quoting ``Huck et al., to be published''. However, this article was only published two years later \cite{1985Huc01}.

\subsubsection*{$^{55,56}$K}

$^{55}$K and $^{56}$K were discovered by Tarasov et al. in 2009 and published in ``Evidence for a change in the nuclear mass surface with the discovery of the most neutron-rich nuclei with 17 $\le$ Z $\le$ 25'' \cite{2009Tar01}. $^9$Be targets were bombarded with 132 MeV/u $^{76}$Ge ions accelerated by the Coupled Cyclotron Facility at the National Superconducting Cyclotron Laboratory at Michigan State University. $^{55}$K and $^{56}$K were produced in projectile fragmentation reactions and identified with a two-stage separator consisting of the A1900 fragment separator and the S800 analysis beam line. ``The observed fragments include fifteen new isotopes that are the most neutron-rich nuclides of the elements chlorine to manganese ($^{50}$Cl, $^{53}$Ar, $^{55,56}$K, $^{57,58}$Ca, $^{59,60,61}$Sc, $^{62,63}$Ti, $^{65,66}$V, $^{68}$Cr, $^{70}$Mn).''

\section{Summary}
The discoveries of the known isotopes and unbound resonances of the elements from sodium to potassium have been compiled and the methods of their production discussed. 194 isotopes were described including 21 stable, 44 proton-rich, 126 neutron-rich, and 3 proton-unbound resonances. Overall the discovery of these isotopes was straightforward. Only for two isotopes ($^{20}$Mg and $^{22}$Mg) the initially measured half-life was incorrect. The half-life of seven isotopes was first reported without a mass assignment ($^{25}$Na, $^{23}$Al, $^{34}$P, $^{38}$Cl, $^{37,42,44}$K). In addition, the half-life of $^{29}$Al was first assigned to $^{27}$Si.

\ack

I would like to thank Ute Thoennessen for carefully proofreading the manuscript. This work was supported by the National Science Foundation under grant No. PHY06-06007 (NSCL).

%%% Here we use thebibliography environment to produce the reference list,
%%% but you can use BibTeX as well:
\bibliography{../isotope-discovery-references}

\newpage

%%% Please start a new page by uncommenting the next
\newpage

\TableExplanation

\bigskip
\renewcommand{\arraystretch}{1.0}

\section*{Table 1.\label{tbl1te} Discovery of isotopes with 11 $\le$ Z $\le$ 19}
\begin{tabular*}{0.95\textwidth}{@{}@{\extracolsep{\fill}}lp{5.5in}@{}}
\multicolumn{2}{p{0.95\textwidth}}{  }\\

Isotope &  Name of isotope \\
First Author & First author of refereed publication \\
Journal & Journal of publication \\
Ref. &  Reference  \\
Method & Production method used in the discovery: \\
    & AS: atomic spectroscopy \\
    & MS: mass spectroscopy \\
    & DI: deep inelastic reactions \\
    & LP: light-particle reactions (including neutrons) \\
    & PF: projectile fragmentation \\
    & PI: pion-induced reactions \\
    & SB: reactions with secondary beams \\
    & SP: spallation reactions \\
    & NC: neutron-capture reactions \\
    & PN: photo-nuclear reactions \\

Laboratory &  Laboratory where the experiment was performed\\
Country &  Country of laboratory\\
Year & Year of discovery  \\
\end{tabular*}
\label{tableI}

\datatables % This command is necessary to get the table names in toc

%% One-page data tables are also best formatted using the longtable
%% environment:
%\begin{longtable}{c}
%\caption{This is the First Data Table}\\
%\endhead\\
%\end{longtable}

%% If the table is to span over the whole text width, we set:

\setlength{\LTleft}{0pt}
\setlength{\LTright}{0pt}

% To avoid ``Overfull \hboxes...'' decrease the intercolumn spacing:

\setlength{\tabcolsep}{0.5\tabcolsep}

\renewcommand{\arraystretch}{1.0}

\footnotesize % we need to squeeze the font size a lot!

\begin{longtable}{@{\extracolsep\fill}rllrllll@{}}
\caption{Discovery of Isotopes with 11 $\le$ Z $\le$ 19. See page\ \pageref{tbl1te} for Explanation of Tables}
Isotope & First Author & Journal & Ref. & Method & Laboratory & Country & Year\\
\hline\\
\endfirsthead\\
\caption[]{(continued)}
Isotope & Author & Journal & Ref. & Method & Laboratory & Country & Year\\
\hline\\
\endhead
$^{18}$Na& T. Zerguerras & Eur. Phys. J. A &\cite{2004Zer01}& SB & GANIL & France &2004 \\
$^{19}$Na& J. Cerny & Phys. Rev. Lett. &\cite{1969Cer01}& LP & Berkeley & USA &1969 \\
$^{20}$Na& L.W. Alvarez & Phys. Rev. &\cite{1950Alv01}& LP & Berkeley & USA &1950 \\
$^{21}$Na& E. Pollard & Phys. Rev. &\cite{1940Pol01}& LP & Yale & USA &1940 \\
$^{22}$Na& O.R. Frisch & Nature &\cite{1935Fri01}& LP & Copenhagen & Denmark &1935 \\
$^{23}$Na& F.W. Aston & Nature &\cite{1921Ast03}& MS & Cambridge & UK &1921 \\
$^{24}$Na& E. Fermi & Proc. Roy. Soc. A &\cite{1934Fer01}& LP & Rome & Italy &1934 \\
$^{25}$Na& O. Huber & Helv. Phys. Acta &\cite{1943Hub02}& PN & Zurich & Switzerland &1943 \\
$^{26}$Na& M.J. Nurmia & Nucl. Phys. A &\cite{1958Nur02}& LP & Arkansas & USA &1958 \\
$^{27}$Na& R. Klapisch & Phys. Rev. Lett. &\cite{1968Kla01}& SP & CERN & Switzerland &1968 \\
$^{28}$Na& R. Klapisch & Phys. Rev. Lett. &\cite{1969Kla01}& SP & CERN & Switzerland &1969 \\
$^{29}$Na& R. Klapisch & Phys. Rev. Lett. &\cite{1969Kla01}& SP & CERN & Switzerland &1969 \\
$^{30}$Na& R. Klapisch & Phys. Rev. Lett. &\cite{1969Kla01}& SP & CERN & Switzerland &1969 \\
$^{31}$Na& R. Klapisch & Phys. Rev. Lett. &\cite{1969Kla01}& SP & CERN & Switzerland &1969 \\
$^{32}$Na& R. Klapisch & Phys. Rev. Lett. &\cite{1972Kla01}& SP & CERN & Switzerland &1972 \\
$^{33}$Na& R. Klapisch & Phys. Rev. Lett. &\cite{1972Kla01}& SP & CERN & Switzerland &1972 \\
$^{34}$Na& M. Langevin & Phys. Lett. B &\cite{1983Lan02}& SP & CERN & Switzerland &1983 \\
$^{35}$Na& M. Langevin & Phys. Lett. B &\cite{1983Lan02}& SP & CERN & Switzerland &1983 \\
$^{36}$Na&   not observed          &               &                &    &      &             & \\
$^{37}$Na& M. Notani & Phys. Lett. B &\cite{2002Not01}& PF & RIKEN & Japan &2002 \\
        &            &               &                &    &       &       &     \\
        &            &               &                &    &       &       &     \\
$^{19}$Mg& I. Mukha & Phys. Rev. Lett. &\cite{2007Muk01}& SB & Darmstadt & Germany &2007 \\
$^{20}$Mg& R.G.H. Robertson & Phys. Rev. Lett. &\cite{1974Rob01}& LP & Juelich & Germany &1974 \\
$^{21}$Mg& R. Barton & Can. J. Phys. &\cite{1963Bar01}& LP & McGill & Canada &1963 \\
$^{22}$Mg& F. Ajzenberg-Selove & Phys. Rev. &\cite{1961Ajz01}& LP & Los Alamos & USA &1961 \\
$^{23}$Mg& M.G. White & Phys. Rev. &\cite{1939Whi01}& LP & Princeton & USA &1939 \\
$^{24}$Mg& A.J. Dempster & Science &\cite{1920Dem01}& MS & Chicago & USA &1920 \\
$^{25}$Mg& A.J. Dempster & Science &\cite{1920Dem01}& MS & Chicago & USA &1920 \\
$^{26}$Mg& A.J. Dempster & Science &\cite{1920Dem01}& MS & Chicago & USA &1920 \\
$^{27}$Mg& E. Fermi & Proc. Roy. Soc. A &\cite{1934Fer01}& LP & Rome & Italy &1934 \\
$^{28}$Mg& R.K. Sheline& Phys. Rev. &\cite{1953She01}& LP & Chicago & USA &1953 \\
$^{29}$Mg& A.G. Artukh & Nucl. Phys. A &\cite{1971Art01}& DI & Dubna & Russia &1971 \\
$^{30}$Mg& A.G. Artukh & Nucl. Phys. A &\cite{1971Art01}& DI & Dubna & Russia &1971 \\
$^{31}$Mg& G.W. Butler & Phys. Rev. Lett. &\cite{1977But01}& SP & Los Alamos & USA &1977 \\
$^{32}$Mg& G.W. Butler & Phys. Rev. Lett. &\cite{1977But01}& SP & Los Alamos & USA &1977 \\
$^{33}$Mg& G.D. Westfall & Phys. Rev. Lett. &\cite{1979Wes01}& PF & Berkeley & USA &1979 \\
$^{34}$Mg& G.D. Westfall & Phys. Rev. Lett. &\cite{1979Wes01}& PF & Berkeley & USA &1979 \\
$^{35}$Mg& D. Guillemaud-Mueller & Z. Phys. A &\cite{1989Gui01}& PF & GANIL & France &1989 \\
$^{36}$Mg& D. Guillemaud-Mueller & Z. Phys. A &\cite{1989Gui01}& PF & GANIL & France &1989 \\
$^{37}$Mg& H. Sakurai & Phys. Rev. C &\cite{1996Sak01}& PF & RIKEN & Japan &1996 \\
$^{38}$Mg& M. Notani & Phys. Lett. B &\cite{2002Not01}& PF & RIKEN & Japan &2002 \\
$^{39}$Mg& not observed &&&&&& \\
$^{40}$Mg& T. Baumann & Nature &\cite{2007Bau01}& PF & Michigan State & USA &2007 \\
        &            &               &                &    &       &       &     \\
        &            &               &                &    &       &       &     \\
$^{22}$Al& M.D. Cable & Phys. Rev. C &\cite{1982Cab01}& LP & Berkeley & USA &1982 \\
$^{23}$Al& J. Cerny & Phys. Rev. Lett. &\cite{1969Cer01}& LP & Berkeley & USA &1969 \\
$^{24}$Al& N.W. Glass & Phys. Rev. &\cite{1953Gla01}& LP & UCLA & USA &1953 \\
$^{25}$Al& J.L.W. Churchill & Nature &\cite{1953Chu01}& LP & Aldermaston & UK &1953 \\
$^{26}$Al& O.R. Frisch & Nature &\cite{1934Fri01}& LP & London & UK &1934 \\
$^{27}$Al& F.W. Aston & Nature &\cite{1922Ast03}& MS & Cambridge & UK &1922 \\
$^{28}$Al& I. Curie & J. Phys. Radium &\cite{1934Cur02}& LP & Paris & France &1934 \\
$^{29}$Al& H.A. Bethe & Phys. Rev. &\cite{1939Bet01}& LP & Purdue & USA &1939 \\
$^{30}$Al& E.L. Robinson & Phys. Rev. &\cite{1961Rob01}& LP & Purdue & USA &1961 \\
$^{31}$Al& A.G. Artukh & Nucl. Phys. A &\cite{1971Art01}& DI & Dubna & Russia &1971 \\
$^{32}$Al& A.G. Artukh & Nucl. Phys. A &\cite{1971Art01}& DI & Dubna & Russia &1971 \\
$^{33}$Al& A.G. Artukh & Nucl. Phys. A &\cite{1971Art01}& DI & Dubna & Russia &1971 \\
$^{34}$Al& G.W. Butler & Phys. Rev. Lett. &\cite{1977But01}& SP & Los Alamos & USA &1977 \\
$^{35}$Al& T.J.M. Symons & Phys. Rev. Lett. &\cite{1979Sym01}& PF & Berkeley & USA &1979 \\
$^{36}$Al& G.D. Westfall & Phys. Rev. Lett. &\cite{1979Wes01}& PF & Berkeley & USA &1979 \\
$^{37}$Al& G.D. Westfall & Phys. Rev. Lett. &\cite{1979Wes01}& PF & Berkeley & USA &1979 \\
$^{38}$Al& D. Guillemaud-Mueller & Z. Phys. A &\cite{1989Gui01}& PF & GANIL & France &1989 \\
$^{39}$Al& D. Guillemaud-Mueller & Z. Phys. A &\cite{1989Gui01}& PF & GANIL & France &1989 \\
$^{40}$Al& M. Notani & Phys. Lett. B &\cite{2002Not01}& PF & RIKEN & Japan &2002 \\
$^{41}$Al& M. Notani & Phys. Lett. B &\cite{2002Not01}& PF & RIKEN & Japan &2002 \\
$^{42}$Al& T. Baumann & Nature &\cite{2007Bau01}& PF & Michigan State & USA &2007 \\
$^{43}$Al& T. Baumann & Nature &\cite{2007Bau01}& PF & Michigan State & USA &2007 \\
        &            &               &                &    &       &       &     \\
        &            &               &                &    &       &       &     \\
$^{22}$Si& M.G.Saint-Laurent & Phys. Rev. Lett. &\cite{1987Sai01}& PF & GANIL & France &1987 \\
$^{23}$Si& M. Langevin & Nucl. Phys. A &\cite{1986Lan01}& PF & GANIL & France &1986 \\
$^{24}$Si& J. \"Ayst\"o & Phys. Lett. B &\cite{1979Ays01}& LP & Berkeley & USA &1979 \\
$^{25}$Si& R. Barton & Can. J. Phys. &\cite{1963Bar01}& LP & McGill & Canada &1963 \\
$^{26}$Si& E.L. Robinson & Phys. Rev. &\cite{1960Rob01}& LP & Purdue & USA &1960 \\
$^{27}$Si& G. Kuerti & Phys. Rev. &\cite{1939Kue01}& LP & Rochester & USA &1939 \\
$^{28}$Si& F.W. Aston & Nature &\cite{1920Ast02}& MS & Cambridge & UK &1920 \\
$^{29}$Si& F.W. Aston & Nature &\cite{1920Ast02}& MS & Cambridge & UK &1920 \\
$^{30}$Si& R.S. Mulliken & Nature &\cite{1924Mul01}& AS & Harvard & USA &1924 \\
$^{31}$Si& E. Fermi & Proc. Roy. Soc. A &\cite{1934Fer01}& LP & Rome & Italy &1934 \\
$^{32}$Si& M. Lindner & Phys. Rev. &\cite{1953Lin01}& LP & Berkeley & USA &1953 \\
$^{33}$Si& A.G. Artukh & Nucl. Phys. A &\cite{1971Art01}& DI & Dubna & Russia &1971 \\
$^{34}$Si& A.G. Artukh & Nucl. Phys. A &\cite{1971Art01}& DI & Dubna & Russia &1971 \\
$^{35}$Si& A.G. Artukh & Nucl. Phys. A &\cite{1971Art01}& DI & Dubna & Russia &1971 \\
$^{36}$Si& A.G. Artukh & Nucl. Phys. A &\cite{1971Art01}& DI & Dubna & Russia &1971 \\
$^{37}$Si& P. Auger & Z. Phys. A &\cite{1979Aug01}& DI & Orsay & France &1979 \\
$^{38}$Si& G.D. Westfall & Phys. Rev. Lett. &\cite{1979Wes01}& PF & Berkeley & USA &1979 \\
$^{39}$Si& G.D. Westfall & Phys. Rev. Lett. &\cite{1979Wes01}& PF & Berkeley & USA &1979 \\
$^{40}$Si& D. Guillemaud-Mueller & Z. Phys. A &\cite{1989Gui01}& PF & GANIL & France &1989 \\
$^{41}$Si& D. Guillemaud-Mueller & Z. Phys. A &\cite{1989Gui01}& PF & GANIL & France &1989 \\
$^{42}$Si& M. Lewitowicz & Z. Phys. A &\cite{1990Lew01}& PF & GANIL & France &1990 \\
$^{43}$Si& M. Notani & Phys. Lett. B &\cite{2002Not01}& PF & RIKEN & Japan &2002 \\
$^{44}$Si& O.B. Tarasov & Phys. Rev. C &\cite{2007Tar01}& PF & Michigan State & USA &2007 \\
        &            &               &                &    &       &       &     \\
        &            &               &                &    &       &       &     \\
$^{26}$P& M.D. Cable & Phys. Lett. B &\cite{1983Cab01}& LP & Berkeley & USA &1983 \\
$^{27}$P& W. Benenson & Phys. Rev. C &\cite{1977Ben01}& LP & Michigan State & USA &1977 \\
$^{28}$P& N.W. Glass & Phys. Rev. &\cite{1953Gla01}& LP & UCLA & USA &1953 \\
$^{29}$P& M.G. White & Phys. Rev. &\cite{1941Whi01}& LP & Princeton & USA &1941 \\
$^{30}$P& I. Curie & Compt. Rend. Acad. Sci. &\cite{1934Cur01}& LP & Paris & France &1934 \\
$^{31}$P& F.W. Aston & Nature &\cite{1920Ast02}& MS & Cambridge & UK &1920 \\
$^{32}$P& E. Fermi & Proc. Roy. Soc. A &\cite{1934Fer01}& LP & Rome & Italy &1934 \\
$^{33}$P& R.K. Sheline& Phys. Rev. &\cite{1951She01}& LP & Chicago & USA &1951 \\
$^{34}$P& W. Z\"unti & Helv. Phys. Acta &\cite{1945Zun01}& LP & Zurich & Switzerland &1945 \\
$^{35}$P& A.G. Artukh & Nucl. Phys. A &\cite{1971Art01}& DI & Dubna & Russia &1971 \\
$^{36}$P& A.G. Artukh & Nucl. Phys. A &\cite{1971Art01}& DI & Dubna & Russia &1971 \\
$^{37}$P& A.G. Artukh & Nucl. Phys. A &\cite{1971Art01}& DI & Dubna & Russia &1971 \\
$^{38}$P& A.G. Artukh & Nucl. Phys. A &\cite{1971Art01}& DI & Dubna & Russia &1971 \\
$^{39}$P& G.W. Butler & Phys. Rev. Lett. &\cite{1977But01}& SP & Los Alamos & USA &1977 \\
$^{40}$P& P. Auger & Z. Phys. A &\cite{1979Aug01}& DI & Orsay & France &1979 \\
$^{41}$P& G.D. Westfall & Phys. Rev. Lett. &\cite{1979Wes01}& PF & Berkeley & USA &1979 \\
$^{42}$P& G.D. Westfall & Phys. Rev. Lett. &\cite{1979Wes01}& PF & Berkeley & USA &1979 \\
$^{43}$P& D. Guillemaud-Mueller & Z. Phys. A &\cite{1989Gui01}& PF & GANIL & France &1989 \\
$^{44}$P& D. Guillemaud-Mueller & Z. Phys. A &\cite{1989Gui01}& PF & GANIL & France &1989 \\
$^{45}$P& M. Lewitowicz & Z. Phys. A &\cite{1990Lew01}& PF & GANIL & France &1990 \\
$^{46}$P& M. Lewitowicz & Z. Phys. A &\cite{1990Lew01}& PF & GANIL & France &1990 \\
        &            &               &                &    &       &       &     \\
        &            &               &                &    &       &       &     \\
$^{27}$S& M. Langevin & Nucl. Phys. A &\cite{1986Lan01}& PF & GANIL & France &1986 \\
$^{28}$S& C.L. Morris & Phys. Rev. C &\cite{1982Mor01}& PI & Los Alamos & USA &1982 \\
$^{29}$S& J.C. Hardy & Phys. Lett. &\cite{1964Har02}& LP & McGill & Canada &1964 \\
$^{30}$S& E.L. Robinson & Phys. Rev. &\cite{1961Rob02}& LP & Purdue & USA &1961 \\
$^{31}$S& L.D.P. King & Phys. Rev. &\cite{1940Kin01}& LP & Purdue & USA &1940 \\
$^{32}$S& F.W. Aston & Nature &\cite{1920Ast02}& MS & Cambridge & UK &1920 \\
$^{33}$S& F.W. Aston & Nature &\cite{1926Ast01}& MS & Cambridge & UK &1926 \\
$^{34}$S& F.W. Aston & Nature &\cite{1926Ast01}& MS & Cambridge & UK &1926 \\
$^{35}$S& E.B. Andersen & Z. Phys. Chemie &\cite{1936And01}& LP & Aarhus & Denmark &1936 \\
$^{36}$S& A.O. Nier & Phys. Rev. &\cite{1938Nie01}& MS & Harvard & USA &1938 \\
$^{37}$S& W. Z\"unti & Helv. Phys. Acta &\cite{1945Zun01}& LP & Zurich & Switzerland &1945 \\
$^{38}$S& D.R. Nethaway & Phys. Rev. &\cite{1958Net01} & LP & Berkeley & USA &1958 \\
$^{39}$S& A.G. Artukh & Nucl. Phys. A &\cite{1971Art01}& DI & Dubna & Russia &1971 \\
$^{40}$S& A.G. Artukh & Nucl. Phys. A &\cite{1971Art01}& DI & Dubna & Russia &1971 \\
$^{41}$S& P. Auger & Z. Phys. A &\cite{1979Aug01}& DI & Orsay & France &1979 \\
$^{42}$S& P. Auger & Z. Phys. A &\cite{1979Aug01}& DI & Orsay & France &1979 \\
$^{43}$S& G.D. Westfall & Phys. Rev. Lett. &\cite{1979Wes01}& PF & Berkeley & USA &1979 \\
$^{44}$S& G.D. Westfall & Phys. Rev. Lett. &\cite{1979Wes01}& PF & Berkeley & USA &1979 \\
$^{45}$S& D. Guillemaud-Mueller & Z. Phys. A &\cite{1989Gui01}& PF & GANIL & France &1989 \\
$^{46}$S& D. Guillemaud-Mueller & Z. Phys. A &\cite{1989Gui01}& PF & GANIL & France &1989 \\
$^{47}$S& D. Guillemaud-Mueller & Z. Phys. A &\cite{1989Gui01}& PF & GANIL & France &1989 \\
$^{48}$S& M. Lewitowicz & Z. Phys. A &\cite{1990Lew01}& PF & GANIL & France &1990 \\
        &            &               &                &    &       &       &     \\
        &            &               &                &    &       &       &     \\
$^{31}$Cl& W. Benenson & Phys. Rev. C &\cite{1977Ben01}& LP & Michigan State & USA &1977 \\
$^{32}$Cl& N.W. Glass & Phys. Rev. &\cite{1953Gla01}& LP & UCLA & USA &1953 \\
$^{33}$Cl& J.B. Hoag & Phys. Rev. &\cite{1940Hoa01}& LP & Berkeley & USA &1940 \\
$^{34}$Cl& O.R. Frisch & Nature &\cite{1934Fri01}& LP & London & UK &1934 \\
$^{35}$Cl& F.W. Aston & Nature &\cite{1919Ast01}& MS & Cambridge & USA &1919 \\
$^{36}$Cl& D.C. Grahame & Phys. Rev. &\cite{1941Gra01}& NC & Berkeley & USA &1941 \\
$^{37}$Cl& F.W. Aston & Nature &\cite{1919Ast01}& MS & Cambridge & USA &1919 \\
$^{38}$Cl& J.W. Kennedy & Phys. Rev. &\cite{1940Ken01}& NC & Berkeley & USA &1940 \\
$^{39}$Cl& R.N.H. Haslam & Phys. Rev. &\cite{1949Has01}& PN & Saskatoon & Canada &1949 \\
$^{40}$Cl& H. Morinaga & Phys. Rev. &\cite{1956Mor01}& LP & Purdue & USA &1956 \\
$^{41}$Cl& A.G. Artukh & Nucl. Phys. A &\cite{1971Art01}& DI & Dubna & Russia &1971 \\
$^{42}$Cl& A.G. Artukh & Nucl. Phys. A &\cite{1971Art01}& DI & Dubna & Russia &1971 \\
$^{43}$Cl& E. Kashy & Phys. Rev. C &\cite{1976Kas01}& LP & Michigan State & USA &1976 \\
$^{44}$Cl& G.D. Westfall & Phys. Rev. Lett. &\cite{1979Wes01}& PF & Berkeley & USA &1979 \\
$^{45}$Cl& G.D. Westfall & Phys. Rev. Lett. &\cite{1979Wes01}& PF & Berkeley & USA &1979 \\
$^{46}$Cl& D. Guillemaud-Mueller & Z. Phys. A &\cite{1989Gui01}& PF & GANIL & France &1989 \\
$^{47}$Cl& D. Guillemaud-Mueller & Z. Phys. A &\cite{1989Gui01}& PF & GANIL & France &1989 \\
$^{48}$Cl& D. Guillemaud-Mueller & Z. Phys. A &\cite{1989Gui01}& PF & GANIL & France &1989 \\
$^{49}$Cl& D. Guillemaud-Mueller & Z. Phys. A &\cite{1989Gui01}& PF & GANIL & France &1989 \\
$^{50}$Cl& O.B. Tarasov & Phys. Rev. Lett. &\cite{2009Tar01}& PF & Michigan State & USA &2009 \\
$^{51}$Cl& M. Lewitowicz & Z. Phys. A &\cite{1990Lew01}& PF & GANIL & France &1990 \\
        &            &               &                &    &       &       &     \\
        &            &               &                &    &       &       &     \\
$^{31}$Ar & M. Langevin & Nucl. Phys. A &\cite{1986Lan01}& PF & GANIL & France &1986 \\
$^{32}$Ar & E. Hagberg & Phys. Rev. Lett. &\cite{1977Hag01}& SP & CERN & Switzerland &1977 \\
$^{33}$Ar & P. L. Reeder & Phys. Rev. Lett. &\cite{1964Ree01}& LP & Brookhaven & USA &1964 \\
$^{34}$Ar & R.G. Miller & Phys. Lett. &\cite{1966Mil01}& LP & Caltech& USA &1966 \\
$^{35}$Ar & L.D.P. King & Phys. Rev. &\cite{1940Kin01}& LP & Purdue & USA &1940 \\
$^{36}$Ar & F.W. Aston & Nature &\cite{1920Ast01}& MS & Cambridge & UK &1920 \\
$^{37}$Ar & P.K. Weimer & Phys. Rev. &\cite{1941Wei01}& LP & Ohio State & USA &1941 \\
$^{38}$Ar & P. Zeeman & Proc. Akad. Soc. &\cite{1934Zee01}& MS & Amsterdam & Netherlands &1934 \\
$^{39}$Ar & A.R. Brosi & Phys. Rev. &\cite{1950Bro01}& LP & Oak Ridge & USA &1950 \\
$^{40}$Ar & F.W. Aston & Nature &\cite{1920Ast01}& MS & Cambridge & UK &1920 \\
$^{41}$Ar & A.H. Snell & Phys. Rev. &\cite{1936Sne01}& LP & Berkeley & USA &1936 \\
$^{42}$Ar & S. Katcoff & Phys. Rev. &\cite{1952Kat01}& NC & Brookhaven & USA &1952 \\
$^{43}$Ar & P.G. Hansen & Phys. Lett. B &\cite{1969Han01}& SP & CERN & Switzerland &1969 \\
$^{44}$Ar & R.E. Larson & Nucl. Phys. A &\cite{1969Lar01}& PN & Naval Research Laboratory & USA &1969 \\
$^{45}$Ar & N.A. Jelley & Phys. Rev. C &\cite{1974Jel01}& LP & Berkeley & USA &1974 \\
$^{46}$Ar & N.A. Jelley & Phys. Rev. C &\cite{1974Jel01}& LP & Berkeley & USA &1974 \\
$^{47}$Ar & D. Guillemaud-Mueller & Z. Phys. A &\cite{1985Gui01}& PF & GANIL & France &1985 \\
$^{48}$Ar & S. Gr\'evy & Phys. Lett. B &\cite{2004Gre01}& PF & GANIL & France &2004 \\
$^{49}$Ar & D. Guillemaud-Mueller & Z. Phys. A &\cite{1989Gui01}& PF & GANIL & France &1989 \\
$^{50}$Ar & D. Guillemaud-Mueller & Z. Phys. A &\cite{1989Gui01}& PF & GANIL & France &1989 \\
$^{51}$Ar & D. Guillemaud-Mueller & Z. Phys. A &\cite{1989Gui01}& PF & GANIL & France &1989 \\
$^{52}$Ar & O.B. Tarasov & Phys. Rev. Lett. &\cite{2009Tar01}& PF & Michigan State & USA &2009 \\
$^{53}$Ar & O.B. Tarasov & Phys. Rev. Lett. &\cite{2009Tar01}& PF & Michigan State & USA &2009 \\
        &            &               &                &    &       &       &     \\
        &            &               &                &    &       &       &     \\
$^{35}$K & W. Benenson & Phys. Rev. C &\cite{1976Ben01}& LP & Michigan State & USA &1976 \\
$^{36}$K & R.E. Berg & Phys. Rev. &\cite{1967Ber01}& LP & Michigan State & USA &1967 \\
$^{37}$K & C.R. Sun & Phys. Rev. &\cite{1958Sun01}& LP & UCLA & USA &1958 \\
$^{38}$K & D.G. Hurst & Phys. Rev. &\cite{1937Hur01}& LP & Berkeley & USA &1937 \\
$^{39}$K & F.W. Aston & Nature &\cite{1921Ast03}& MS & Cambridge & UK &1921 \\
$^{40}$K & A.O. Nier & Phys. Rev. &\cite{1935Nie01}& MS & Minnesota & USA &1935 \\
$^{41}$K & F.W. Aston & Nature &\cite{1921Ast03}& MS & Cambridge & UK &1921 \\
$^{42}$K & G. Hevesy & Nature &\cite{1935Hev03}& LP & Copenhagen & Denmark &1935 \\
$^{43}$K & R. Overstreet & Phys. Rev. &\cite{1949Ove01}& LP & Berkeley & USA &1949 \\
$^{44}$K & B.L. Cohen & Phys. Rev. &\cite{1954Coh02}& LP & Oak Ridge & USA &1954 \\
$^{45}$K & H. Morinaga & Phys. Lett. &\cite{1964Mor01}& LP & Amsterdam & Netherlands &1964 \\
$^{46}$K & A. Marinov & Phys. Lett. &\cite{1965Mar01}& LP & Argonne & USA &1965 \\
$^{47}$K & T. Kuroyanagi & Nucl. Phys. &\cite{1964Kur01} & PN & JAERI & Japan &1964 \\
$^{48}$K & R. Klapisch & Phys. Rev. Lett. &\cite{1972Kla01}& SP & CERN & Switzerland &1972 \\
$^{49}$K & R. Klapisch & Phys. Rev. Lett. &\cite{1972Kla01}& SP & CERN & Switzerland &1972 \\
$^{50}$K & R. Klapisch & Phys. Rev. Lett. &\cite{1972Kla01}& SP & CERN & Switzerland &1972 \\
$^{51}$K & M. Langevin & Phys. Lett. B &\cite{1983Lan01}& SP & CERN & Switzerland &1983 \\
$^{52}$K & M. Langevin & Phys. Lett. B &\cite{1983Lan01}& SP & CERN & Switzerland &1983 \\
$^{53}$K & M. Langevin & Phys. Lett. B &\cite{1983Lan01}& SP & CERN & Switzerland &1983 \\
$^{54}$K & M. Langevin & Phys. Lett. B &\cite{1983Lan01}& SP & CERN & Switzerland &1983 \\
$^{55}$K & O.B. Tarasov & Phys. Rev. Lett. &\cite{2009Tar01}& PF & Michigan State & USA &2009 \\
$^{56}$K & O.B. Tarasov & Phys. Rev. Lett. &\cite{2009Tar01}& PF & Michigan State & USA &2009 \\
\\
\end{longtable}

\end{document}